\documentclass[fleqn,usenatbib]{mnras}
\usepackage{amsmath}
\usepackage{newtxtext,newtxmath}
\usepackage[dvipsnames]{xcolor}

\usepackage[T1]{fontenc}

\DeclareRobustCommand{\VAN}[3]{#2}
\let\VANthebibliography\thebibliography
\def\thebibliography{\DeclareRobustCommand{\VAN}[3]{##3}\VANthebibliography}

\usepackage{graphicx}	% Including figure files
\usepackage{amsmath}	% Advanced maths commands
\title[Short title, max. 45 characters]{What Determines the Boundaries of H$_{2}$O Maser Emission in an X-ray Illuminated Gas Disk ?}

\author[C. Y. Kuo et al.]{
C. Y. Kuo,$^{1,2}$\thanks{E-mail: cykuo.tara@g-mail.nsysu.edu.tw (NSYSU)}
F. Gao,$^{3}$
J. A. Braatz,$^{4}$
D. W. Pesce,$^{5,6}$
E. M. L. Humphreys,$^{7,8}$
M. J. Reid,$^{5}$
\newauthor
C. M. V. Impellizzeri,$^{9}$
C. Henkel,$^{10}$
J. Wagner,$^{10}$
C. E. Wu,$^{1,11}$
\\
$^{1}$Physics Department, National Sun Yat-Sen University, No. 70, Lien-Hai Rd, Kaosiung City 80424, Taiwan, R.O.C\\
$^{2}$Academia Sinica Institute of Astronomy and Astrophysics, P.O. Box 23-141, Taipei 10617, Taiwan, R.O.C. \\
$^{3}$Hamburg Observatory, Gojenbergsweg 112, 21029 Hamburg\\
$^{4}$National Radio Astronomy Observatory, 520 Edgemont Road, Charlottesville, VA 22903, USA\\
$^{5}$Center for Astrophysics $|$ Harvard \& Smithsonian, 60 Garden Street, Cambridge, MA 02138, USA\\
$^{6}$Black Hole Initiative at Harvard University, 20 Garden Street, Cambridge, MA 02138, USA\\
$^{7}$Joint ALMA Observatory, Alonso de Cordova 3107, Vitacura, Santiago, Chile \\
$^{8}$European Southern Observatory (ESO) Vitacura, Alonso de Cordova 3107, Vitacura, Santiago, Chile\\
$^{9}$Leiden Observatory, Leiden University, PO Box 9513, 2300 RA Leiden, The Netherlands\\
$^{10}$Max-Planck Institut für Radioastronomie, Auf dem Hügel 69, 53121 Bonn, Germany\\
$^{11}$Center for Astronomy, National Tsing-Hua University (NTHU), No. 101, Section 2, Kuang-Fu Road, Hsinchu 30013, Taiwan, R.O.C
}

\date{Accepted XXX. Received YYY; in original form ZZZ}

\pubyear{2015}

\begin{document}
\label{firstpage}
\pagerange{\pageref{firstpage}--\pageref{lastpage}}
\maketitle

\begin{abstract}
High precision mapping of H$_{2}$O megamaser emission from active galaxies has revealed more than a dozen Keplerian H$_{2}$O maser disks, which enable a $\sim$4\% uncertainty estimate of the Hubble constant as well as providing accurate masses for the central black holes. These disks often have well-defined inner and outer boundaries of maser emission on sub-parsec scales. In order to better understand the physical conditions that determine the inner and outer radii of a maser disk, we examine the distributions of gas density and X-ray heating rate in a warped molecular disk described by a power-law surface density profile. For a suitable choice of the disk mass, we find that the outer radius $R_{\rm out}$ of the maser disk predicted from our model can match the observed value, with $R_{\rm out}$ mainly determined by the maximum heating rate or the minimum density for efficient maser action, depending on the combination of the Eddington ratio, black hole mass, and disk mass. Our analysis also indicates that the inner radius for maser action is comparable to the dust sublimation radius, suggesting that dust may play a role in determining the inner radius of a maser disk. Finally, our model predicts that H$_{2}$O gigamaser disks could exist at the centers of high-z quasars, with disk sizes of $\gtrsim 10-30$ pc.  

\end{abstract}


%%%%%%%%%%%%%%%%%%%%%%%%%%%%%%%%%%%%%%%%%%%%%%%%%%

%%%%%%%%%%%%%%%%% BODY OF PAPER %%%%%%%%%%%%%%%%%%

\section{Introduction}
In the nuclear regions of external galaxies hosting active galactic nuclei (AGNs), there exist powerful cosmic masers from the $J_{K_{-}K_{+}}=6_{16}-5_{23}$ transition\footnote{{Here, $J$ is the total angular momentum of the H$_{2}$O molecule, with $K_{-}$ and $K_{+}$ representing the projections of $J$ on two molecular axes \citep[e.g.][]{cooke_elitzur_85}.}} of the ortho-H$_{2}$O molecule emitting at 22.23508 GHz, which arise either from a sub-parsec circumnuclear disk \citep[e.g.][]{kuo11, gao16} or a nuclear wind or jet \citep[e.g.][]{claussen98, gbe03, shoko08, kuo20a, stc20}. These 22 GHz H$_{2}$O {\it megamasers} often display total maser luminosities $\gtrsim$10$^{6}$ greater than that of typical Galactic maser sources, and their extremely high surface brightnesses permit mapping at sub-milliarcsecond resolution using Very Long Baseline interferometry (VLBI), providing a unique probe of the gas distribution and kinematics on sub-parsec scales at the centers of distant galaxies \citep{lo05}.

In the archetypal H$_{2}$O maser galaxy NGC 4258 \citep[e.g.][]{haschick_baan1994, argon07, hum08}, the masing gas resides between radii of $\sim$0.13$-$0.26 pc in a thin disk that is viewed in almost edge-on and follows Keplerian rotation. Such disk maser systems enable black hole (BH) mass measurements to percent-level accuracy \citep[e.g.][]{kuo11, gao17} and provide an accurate geometric distance measurement independent of distance ladders and standard candles \citep[e.g.][]{reid09,gao16,dom20a}. 

To identify disk megamasers like NGC 4258 for measuring the Hubble constant $H_{0}$, the Megamaser Cosmology Project (MCP; \citealt{reid09, bra10}) has carried out an extensive survey of H$_{2}$O megamaser emission from $>$4800 AGNs \citep{kuo18_SED,kuo20b}, resulting in the detection of $\gtrsim$30 candidate disk masers \citep{dom15}. The follow-up imaging of these candidates has increased the number of H$_{2}$O maser disks with high precision VLBI maps by a factor of $\gtrsim$4 over the past decade. This progress not only enables an $H_{0}$ measurement with 4\% uncertainty \citep{dom20b}, but also allows us to explore the black hole-host galaxy coevolution \citep[e.g.][]{kh13, gre16}. In addition, it provides a new tool for constraining the spins of supermassive BHs \citep{masini22}. 

While progress in the detection and mapping of H$_{2}$O maser disks has been significant in the past decade, the physics of H$_{2}$O maser disks has received less attention and is not well-explored. In particular, it is not yet clear what causes the warps seen in many Keplerian maser disks \citep[e.g.][]{tal09, tal12, cap07, kuo11, Kamali19}. In addition, it is still uncertain what determines the inner and outer boundaries of a maser disk and why H$_{2}$O megamaser emissions are typically confined within the narrow range of $\sim$0.1$-$1 pc in a circumnuclear disk \citep[e.g.][]{nm95, gao17, wy12}. Addressing questions like these is valuable because it would not only improve our understanding of the maser disks themselves, but would also allow one to explore unrecognized systematics when using the H$_{2}$O megamaser technique for $H_{0}$ and BH mass measurement. Moreover, a deeper knowledge of maser disk physics can also allow us to investigate whether hyper-luminous H$_{2}$O ``{\it gigamaser}" disks could exist in the high-redshift universe \citep{lo05}. Investigating this possibility is important because such gigamasers, if they exist, would enable the application of the maser technique to high-z galaxies for measuring dynamical BH masses and accurate distances, opening a new avenue to test cosmological models \citep[e.g.][]{king14, lusso19} that might address the ``{\it Hubble tension problem}'' \citep[e.g.][]{val21}.

In this paper, we aim to explore whether the physical conditions favorable for population inversion of H$_{2}$O molecules could play the primary role for determining the inner and outer radii of an H$_{2}$O maser disk.
%, which refers specifically to the section of a gas disk within which H$_{2}$O maser emissions could arise. 
We focus primarily on the physics of the masing region, because this is the part of the disk that is relevant for $H_{0}$ and $M_{\rm BH}$ measurements based on the H$_{2}$O megamaser technique. Understanding the mechanisms that determines the physical size of the masing section would allow us to explore whether the megamaser technique could be applied to galaxies in a redshift range beyond the current limit of the MCP (i.e. $z\lesssim$0.05).  

In \autoref{sec:2}, we discuss measurements of maser disk radii based on VLBI maps compiled from the literature. In \autoref{sec:3}, we make predictions of the inner and outer radii of a maser disk based on an examination of the physical conditions in a circumnuclear disk, and we compare these predictions with the observed values. A discussion of the possibility of applying the maser technique to high-z H$_{2}$O gigamasers is presented in \autoref{sec:4}, and our present study is summarized in \autoref{sec:5}.

\vspace{-0.2 cm}
\begin{table*}
%\small
\caption{The Megamaser Sample, the Inner/Outer Radii and the Warp parameters of the Disks}\label{tab1}
    \centering
    \begin{tabular}{l l l l l l c l l l c}
  \hline\hline       
Galaxy & $M_{\rm BH}$  & $R_{\rm in}$ &  $R_{\rm out}$  & $R_{\rm in}$ &  $R_{\rm out}$  &   $R_{\rm out}$/$R_{\rm in}$ & $\mu_{\rm in}$  & $\mu_{\rm out}$ & $\mu_{\rm in}/\mu_{\rm out}$ & Reference  \\ 
Name   &  ($10^{7}M_{\sun}$) &   (pc)       &    (pc)        &     (10$^{5}$ $R_{s}$)    &   (10$^{5}$ $R_{s}$)         &                &           &       &    \\ 
\hline
  NGC4258   &  4.0$\pm$0.09  & 0.11$^{+0.004}_{-0.004}$   &  0.29$^{+0.01}_{-0.01}$   &  0.29$^{+0.01}_{-0.01}$    &  0.76$^{+0.03}_{-0.03}$  &     2.64  & 0.121$\pm$0.002  & 0.26$\pm$0.005  & 0.465$\pm$0.001  &   1 \\
  NGC1068   &  1.7$\pm$0.02  & 0.58$^{+0.02}_{-0.02}$   &  1.05$^{+0.02}_{-0.02}$   &  3.57$^{+0.12}_{-0.12}$    &  6.46$^{+0.12}_{-0.12}$  &     1.81  & ---  & 0.07$\pm$0.009  & ---  &   2 \\
   NGC2273   &  0.75$\pm$0.05  & 0.03$^{+0.01}_{-0.01}$   &  0.08$^{+0.29}_{-0.01}$   &  0.42$^{+0.14}_{-0.14}$    &  1.12$^{+4.05}_{-0.14}$  &     2.67  & 0.356$\pm$0.013  & 0.337$\pm$0.028  & 1.056$\pm$0.049  &   1 \\
  IC2560   &  0.44$\pm$0.09  & 0.11$^{+0.02}_{-0.02}$   &  0.47$^{+0.03}_{-0.03}$   &  2.62$^{+0.48}_{-0.48}$    &  11.18$^{+0.71}_{-0.71}$  &     4.27  & 0.21$\pm$0.003  & 0.13$\pm$0.005  & 1.615$\pm$0.039  &   1 \\
   UGC3789   &  1.04$\pm$0.06  & 0.08$^{+0.02}_{-0.02}$   &  0.3$^{+0.19}_{-0.03}$   &  0.81$^{+0.2}_{-0.2}$    &  3.02$^{+1.91}_{-0.3}$  &     3.75  & 0.017$\pm$0.005  & 0.345$\pm$0.009  & 0.049$\pm$0.013  &   1 \\
   NGC1194   &  6.5$\pm$0.4  & 0.54$^{+0.03}_{-0.25}$   &  1.33$^{+0.97}_{-0.06}$   &  0.87$^{+0.05}_{-0.4}$    &  2.14$^{+1.56}_{-0.1}$  &     2.46  & 0.075$\pm$0.001  & 0.035$\pm$0.001  & 2.143$\pm$0.033  &   1 \\
   NGC3393   &  3.1$\pm$0.37  & 0.17$^{+0.02}_{-0.02}$   &  0.6$^{+0.76}_{-0.04}$   &  0.57$^{+0.07}_{-0.07}$    &  2.03$^{+2.57}_{-0.14}$  &     3.53  & 0.172$\pm$0.022  & 0.047$\pm$0.016  & 3.660$\pm$0.778  &   1 \\
J0437+2456   &  0.29$\pm$0.03  & 0.06$^{+0.03}_{-0.03}$   &  0.13$^{+0.03}_{-0.03}$   &  2.17$^{+1.08}_{-1.08}$    &  4.69$^{+1.08}_{-1.08}$  &     2.17  & 0.208$\pm$0.061  & 0.248$\pm$0.093  & 0.839$\pm$0.069  &   1 \\
   NGC2960   &  1.16$\pm$0.07  & 0.13$^{+0.04}_{-0.04}$   &  0.37$^{+0.71}_{-0.04}$   &  1.17$^{+0.36}_{-0.36}$    &  3.34$^{+6.41}_{-0.36}$  &     2.85  & 0.054$\pm$0.011  & 0.043$\pm$0.024  & 1.256$\pm$0.445  &   1 \\
   NGC5495   &  1.05$\pm$0.2  & 0.1$^{+0.05}_{-0.05}$   &  0.3$^{+0.05}_{-0.05}$   &  1.0$^{+0.5}_{-0.5}$    &  2.99$^{+0.5}_{-0.5}$  &      3.0  & ---  & 0.136$\pm$0.107  & ---  &   1 \\
 CGCG074-064   &  2.42$\pm$0.2  & 0.12$^{+0.01}_{-0.01}$   & 0.47$^{+0.04}_{-0.04}$   &  0.53$^{+0.05}_{-0.05}$    &  2.04$^{+0.18}_{-0.18}$  &     3.92  & 0.032$\pm$0.004  & 0.037$\pm$0.002  & 0.865$\pm$0.036  &   3 \\
   NGC6323   &  0.94$\pm$0.04  & 0.13$^{+0.05}_{-0.05}$   &  0.3$^{+0.05}_{-0.05}$   &  1.45$^{+0.56}_{-0.56}$    &  3.34$^{+0.56}_{-0.56}$  &     2.31  & 0.108$\pm$0.011  & 0.145$\pm$0.023  & 0.745$\pm$0.042  &   1 \\
ESO558-G009   &  1.7$\pm$0.14  & 0.2$^{+0.05}_{-0.05}$   &  0.47$^{+0.06}_{-0.06}$   &  1.23$^{+0.31}_{-0.31}$    &  2.89$^{+0.37}_{-0.37}$  &     2.35  & 0.055$\pm$0.023  & 0.042$\pm$0.042  & 1.310$\pm$0.762  &   1 \\
  NGC5765b   &  4.55$\pm$0.31  & 0.3$^{+0.06}_{-0.06}$   &  1.15$^{+0.07}_{-0.07}$   &  0.69$^{+0.14}_{-0.14}$    &  2.65$^{+0.16}_{-0.16}$  &     3.83  & 0.039$\pm$0.002  & 0.094$\pm$0.006  & 0.415$\pm$0.005 &   1 \\
   NGC6264   &  2.91$\pm$0.11  & 0.24$^{+0.07}_{-0.07}$   &  0.8$^{+0.07}_{-0.07}$   &  0.86$^{+0.25}_{-0.25}$    &  2.88$^{+0.25}_{-0.25}$  &     3.33  & 0.182$\pm$0.008  & 0.166$\pm$0.023  & 1.096$\pm$0.104  &   1 \\
   UGC6093   &  2.65$\pm$0.23  & 0.12$^{+0.07}_{-0.07}$   &  0.24$^{+0.3}_{-0.08}$   &  0.47$^{+0.28}_{-0.28}$    &  0.95$^{+1.18}_{-0.32}$  &      2.0  & 0.058$\pm$0.008  & 0.038$\pm$0.006  & 1.526$\pm$0.030  &   1 \\

\hline   

    \end{tabular}   
\newline
 \raggedleft 
\footnotesize{{\bf Note.} Column (1): source name listed in the order of increasing galaxy distance; Column (2): BH mass; Columns (3) \& (4): the inner and outer radii in units of parsecs; Columns (5) \& (6): the inner and outer radii in units of $1\times 10^{5}$ Schwarzschild radii ($ R_{\rm S}$); Column (7): the ratio between $R_{\rm out}$ and $R_{\rm in}$; Columns (8) \& (9): the obliquity parameters at the inner ($\mu_{\rm in}$) and outer ($\mu_{\rm out}$) edges of the warped maser disk; Column (10): the ratio between $\mu_{\rm in}$ and $\mu_{\rm out}$; Column (11): the reference from which we obtain the BH mass and the disk size: 1. \citet{gao17}; 2. \citet{gallimore23}; 3. \citet{dom20a}. \newline
}
    \label{tab:1}
\end{table*}

\section{The Physical Size of an H$_{2}$O Maser Disk} \label{sec:2}

\subsection{The Radius Measurement}
H$_{2}$O megamasers in a disk configuration ({\it maser disks} hereafter) often display three distinct spatial and velocity groups, including the systemic, redshifted, and blueshifted maser features \citep{kuo11, dom15}. In nearly all cases, the VLBI maps of such maser disks show well-defined inner and outer boundaries of maser emissions on sub-parsec scales, indicated by the positions of the highest and lowest velocity components of either blueshifted or redshifted masers. When the position of the dynamical center of the disk is determined either from disk modeling \citep[e.g.][]{reid13,kuo13,kuo15} or rotation curve fitting for the BH mass \citep[e.g.][]{kuo11,gao17}, the physical inner ($R_{\rm in}$) and outer ($R_{\rm out}$) radii of a maser disk can be measured {\it directly} from its VLBI map given the distance. Alternatively, if one assumes that the maser emission originates only from a disk, without any components associated with a jet or outflow, one could also infer the physical disk radii {\it indirectly} from a single-dish spectrum of a maser disk based on the velocities of the high-velocity maser features and assuming Keplerian rotation\footnote{{\it The high-velocity maser features} refer to the redshifted and blueshifted maser components of a maser disk \citep{kuo11} } 
about a known central mass
\citep[see details in][]{gao17}. 

For both direct and indirect methods, measuring the physical radii of the inner and outer boundaries of a maser disk from its projected angular radii requires the knowledge of the galaxy distance, which can be obtained either from modeling of the maser disk itself or from published values (e.g. NED; https://ned.ipac.caltech.edu/). Since a galaxy distance from disk modeling is available only for a limited number ($\sim$6) of disk maser systems \citep{dom20b}, we simply adopt the Hubble flow distances from NED\footnote{The Hubble flow distance in NED is evaluated based on the standard $\Lambda$CDM cosmology,
with $\Omega_{\rm matter} = 0.308$, $\Omega_{\rm vacuum} = 0.692$, and $H_{0} = 67.8$ km~s$^{-1}$Mpc$^{-1}$.} to infer the physical sizes for all disk maser systems discussed in this paper for consistency.

To identify the physical mechanism that determines the physical size of an H$_{2}$O maser disk, we first compile in \autoref{tab:1} all H$_{2}$O megamasers from literature that display geometrically thin maser disks in their VLBI maps \citep[e.g.][]{kuo11, reid13, gao17, dom20a}. The majority of the radius measurements are drawn from Table 5 in \citet{gao17}, which provides reliable estimates of disk radii for {\it all} ``clean''\footnote{A ``clean'' maser disk indicates that all maser emissions come from the disk, with no maser components associated with a jet or outflow.} H$_{2}$O maser disks that follow Keplerian rotation.  We exclude NGC 4388 listed in \citet{gao17} from our analysis because the lack of systemic maser features in this system plus the small number of high-velocity maser components ($\sim$4) makes it difficult to be confident that the maser emission originates from a thin, rotating disk. For completeness, we also include the thin maser disk in NGC 1068 in our analysis. This source was first reported to be a complicated maser disk that follows sub-Keplerian rotation \citep{greenhill96}, with a disk mass comparable to the BH mass \citep{lodato2003}. However, the latest VLBI observations and analysis \citep{gallimore23} suggest an opposite interpretation for NGC 1068, showing that the disk kinematics are consistent with Keplerian rotation and low turbulence speeds, with a disk mass $\sim$100 times smaller than $M_{\rm BH}$. Therefore, although NGC 1068 could be more complicated than other Keplerian maser systems, considering that the mechanism determining the maser disk size may not be dependent on gas kinematics alone, we still include it. In total, our sample thus consists of 16 maser disks. The BH masses and the inner/outer disk radii of these systems are shown in Column (2) through (6) in \autoref{tab:1}.

\begin{figure*}
\begin{center} 
%\vspace*{-0.3 cm} 
\hspace*{0.0 cm} 
\includegraphics[angle=0, scale= 0.7]{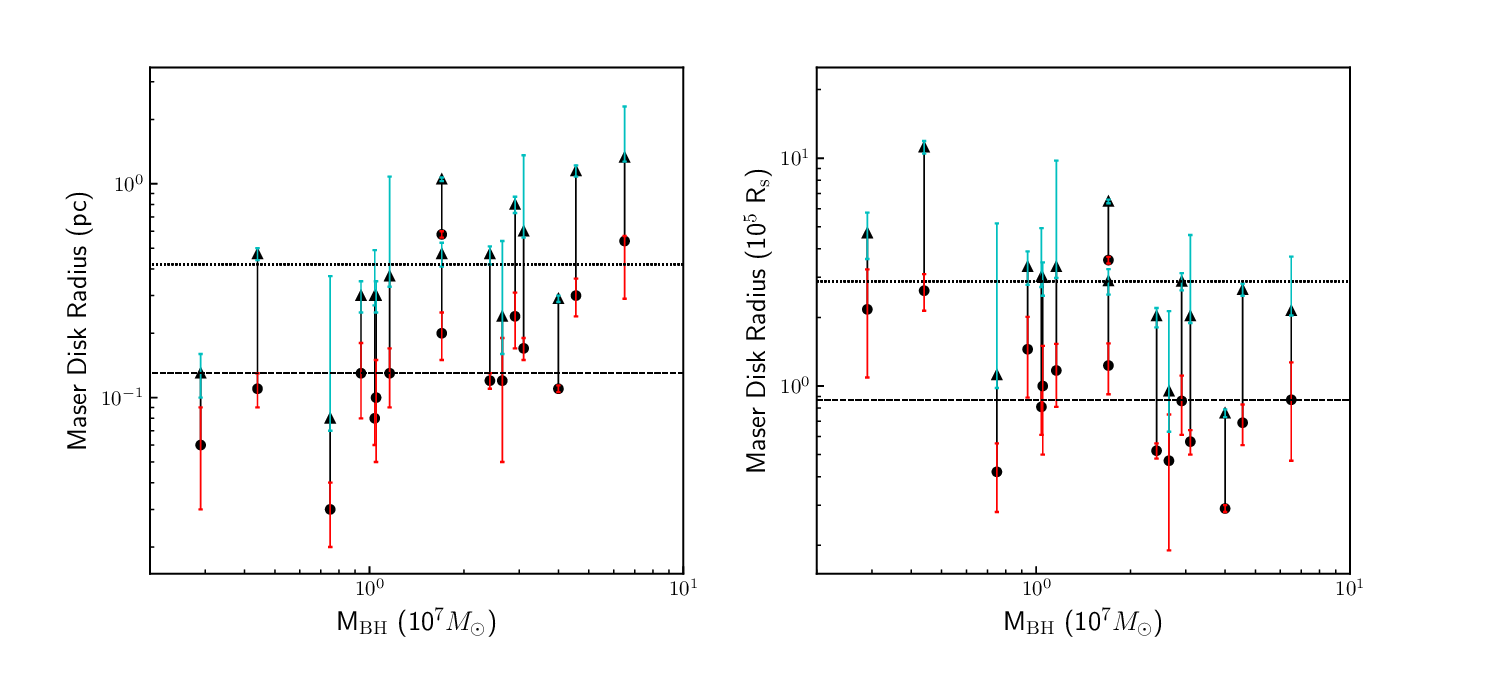} 
\vspace*{-0.5 cm} 
\caption{Left panel: The maser disk size in units of parsecs. The top and bottom of each black vertical bar marked by the filled triangle and black circle indicate the outer ($R_{\rm out}$) and inner ($R_{\rm in}$) radii of a maser disk, respectively, with the blue and red error bars showing the uncertainties in the measurements of $R_{\rm out}$ and $R_{\rm in}$, respectively. The horizontal axis denotes the BH mass for each maser system. The black dashed line and dotted line in the plot indicates the radii of 0.12 pc and 0.42 pc, the medians of $R_{\rm in}$ and $R_{\rm out}$ for all sources, respectively; Right panel: the maser disk size in units of 10$^{5}$ Schwarzschild radii ($R_{\rm S}$). The majority of the disk megamasers in our sample have inner and outer radii within a factor of $\sim$2 from $R_{\rm in} \sim 1\times 10^{5} R_{\rm S}$ and $R_{\rm out} \sim 3\times 10^{5} R_{\rm S}$, respectively. The horizontal dashed line and dotted line indicate the medians of $R_{\rm in}$ ($0.87\times10^{5} R_{\rm S}$) and $R_{\rm out}$ ($2.88\times10^{5} R_{\rm S}$) for all sources.}

\label{fig:1} 
\end{center} 
\end{figure*}

% ****************Remember to use the citation mode of the figure and table************
% -------------------------------------------------------------------------------------
In the left panel of \autoref{fig:1}, we plot disk radii in units of parsec against the BH masses of our maser sample. The top and bottom edges of each black vertical line marked by the filled black circles represent $R_{\rm out}$ and $R_{\rm in}$ for a maser disk, respectively, revealing that the maser emission is mostly confined within the radial range of $\sim 0.1-1$ pc for the majority of the sources. The red and blue error bars in the plot show the uncertainties of $R_{\rm in}$ and $R_{\rm out}$, respectively. In addition to the uncertainties from the galaxy distance and maser position measurements, the error bars also include the discrepancy between the disk radii directly measured from VLBI maps and indirectly inferred from high-sensitivity single-dish maser spectra \citep{dom15}, which could indicate a systematic uncertainty arising from limitations in the spectral coverage or sensitivity in the VLBI observation. 

In the MCP, the single-dish monitoring of the maser galaxies usually covers a $\sim$3 times wider spectral range than the VLBI observation. In the case of NGC 1194, for example, the high-sensitivity single-dish spectrum \citep{dom15} shows that the (weak) redshifted maser feature having the highest velocity lies well outside the spectral coverage of the VLBI observations \citep{kuo11}, suggesting that the true inner radius of the disk may be smaller than the value derived from the VLBI map. In addition, the spectra of some sources (e.g. NGC 3393) sometimes reveal faint, isolated maser features lying between the systemic and the high-velocity maser complexes in the spectra. Without sensitive VLBI mapping, it is difficult to discern whether such features are systemic masers, outflow components, or high-velocity maser features residing close to the outer boundary of the maser disk, leading to uncertainties in $R_{\rm out}$. For a more detailed discussion on analysis of the uncertainties due to spectral coverage, we refer the readers to \citet{gao17}.

\subsection{The Sensitivity Effect on Radius Measurement}

It is well known that the flux densities of the maser emissions usually drop quickly at the wings of the maser line complexes in a Keplerian maser disk \citep[e.g.][]{kuo11,dom15,gao17}. Because of limited sensitivity, one may not necessarily be able to detect the weakest maser features that define the intrinsic size of a maser disk. As a consequence, a VLBI observation will inevitably reveal sharp edges of the maser disk in the maser map, and the observed values of the inner and outer radii of the maser disk can vary depending on the sensitivity of the observation.

To estimate the magnitude of the variation in the radius measurement caused by the sensitivity limit, one can consider the most well-known maser disk in NGC 4258, which has a distance of 7.6 Mpc \citep{hum13}, the closest system in our sample. Its proximity allows for the detection of the weakest maser features among all maser systems studied in this work. We note that if NGC 4258 were placed at a distance of 65 Mpc, the median distance of our maser sample \footnote{The source distances are listed in Table \ref{tab:2}.}, the flux densities of the masers would drop by a factor of $\sim$70. Given the same VLBI sensitivity \citep{argon07}, the observed inner and outer radii of NGC 4258 would change by $\sim$45\% and $\sim$12\%, respectively. We expect that these variations give the magnitude of the uncertainty caused by the sensitivity limit in the radius measurement for our sample if one compares the disk sizes at the same median distance. Since this uncertainty is comparable to the radius errors shown in Table 1, which include uncertainties inferred from sensitive single-dish spectra, the conclusions for this work are not significantly affected by the sensitivity-dependent effect on disk size measurement.

\subsection{The Characteristic Radii of H$_{2}$O maser Disks}

As one can see in the left panel of Figure \ref{fig:1}, the mean radius of a maser disk appears to increase with the BH mass. This trend was first reported by \citet{wy12} based on the data of 8 maser disks, showing that the outer radii can be approximately described by $R_{\rm out} = 0.3(M_{\rm BH}/10^{7}M_{\sun}$) pc. With the inclusion of 6 more maser disks in their analysis, \cite{gao17} obtained a significantly different scaling ($R_{\rm out} \propto M_{\rm BH}^{0.57\pm 0.16}$) and show that both $R_{\rm in}$ and $R_{\rm out}$ are well correlated with the BH mass\footnote{The Spearman's rank correlation coefficients are 0.71 and 0.62 for $R_{\rm in}$ and $R_{\rm out}$, respectively \citep{gao17}}, suggesting that $M_{\rm BH}$ plays an important role in determining the inner and outer radii of a maser disk. Because of the significant correlation between $R_{\rm out}$ and $M_{\rm BH}$, we conjecture that the maser disks may reveal an interesting, characteristic scale if the disk size is expressed in units of the Schwarzschild radius $R_{\rm S}=2GM_{\rm BH}/c^{2}$, where G and c are the gravitational constant and the speed of light, respectively.

In the right panel of Figure \ref{fig:1}, we plot the inner and outer radii of the maser disks in units of $10^{5} R_{\rm S}$. As one can see in this plot, the majority of our maser disks have inner and outer radii within a factor of $\sim$2 from $R_{\rm in} \sim 1\times 10^{5}$ $R_{\rm S}$ and $R_{\rm out} \sim 3\times 10^{5}$ $R_{\rm S}$, which are shown by the horizontal dashed and dotted lines, respectively. It is likely that the presence of these characteristic disk radii may be associated with the fine-tuning nature of the H$_{2}$O megamaser phenomenon, and we will explore this in more detail in Section \ref{sec:4.1} based on the disk model presented in the subsequent section.

%As we will show in Section \ref{sec:3.6}, the existence of the characteristic inner radius and the outlier can be understood if the inner edge of a maser disk lies close to the dust sublimation radius of a local Seyfert galaxy.}

% The only more significant outlier is NGC 1068, whose inner and outer radii are $R_{\rm in} \sim7.6\times 10^{5}$ $R_{\rm S}$ and $R_{\rm out} \sim 13.7\times 10^{5}$ $R_{\rm S}$. The available proposals that provide tentative explanations for the maser disk size \citep[i.e.][]{nm95, wy12, gao17} do not give a clear reason why most maser disks have this characteristic size of $\sim$$(1-3)\times 10^{5}$ $R_{\rm S}$. In addition, it is puzzling why NGC 1068 should be such an outlier.

%while the outlier suggests that some physical parameters in addition to $M_{\rm BH}$ may become important for determining the maser disk size in certain circumstances. 

\section{The Mechanisms that Determine the Size of a Maser Disk }\label{sec:3}
\subsection{The Physical Conditions for Maser Pumping}\label{sec:3.1}
In the interstellar medium, the $J_{K_{-}K_{+}}=6_{16}-5_{23}$ water maser transition can occur naturally through collisional pumping in a warm molecular cloud. In the limit where the medium is optically thick to the far-infrared continuum, the water level populations for a saturated maser source\footnote{In the on-going analysis of the H$_{2}$O maser spectra (Kuo et al. in prep.) for disk maser systems listed in Table \ref{tab:1}, we find no concrete evidence for line narrowing \citep[e.g.][]{baudry2023} which would indicate {\it unsaturated} masers in the majority of the systems. Therefore, we make the simple assumption that all maser sources are saturated for the discussion in the rest of the paer.} are mainly determined by the local conditions including the gas temperature ($T_{\rm H_{2}}$), the dust temperature $T_{\rm d}$, the number density ($n_{\rm H_{2}}$) of molecular gas, and the water abundance $x_{\rm H_{2}O}$ $\equiv$ $n_{\rm H_{2}O}$/$n_{\rm H_{2}}$ \citep{neu2000}. Among these local conditions, the most important ones for maser pumping are $T_{\rm H_{2}}$ and $n_{\rm H_{2}}$, which need to fall within the favored ranges of $400 \lesssim T_{\rm H_{2}} \lesssim 1500$ K and $10^{7} \lesssim n_{\rm H_{2}} \lesssim 10^{11}$ cm$^{-3}$, respectively \citep[e.g.][]{neu94, neu2000, herrn05, lo05, gray16}. The minimum gas temperature of $T_{\rm min}\sim400$ K is required for efficient collisional pumping to the $J_{K_{-}K_{+}}=6_{16}$ level lying at E/k = 643 K above the ground level. Moreover, this temperature threshold is also essential to make a significant enhancement of the water abundance in the cloud \citep[$x_{\rm H_{2}O}$ $\gtrsim$10$^{-4}$; e.g.][]{neu2000} with the reaction O$~+~$H$_{2}$ $\rightarrow$ OH$~+~$H followed by OH~$+$~H$_{2}$ $\rightarrow$ H$_{2}$O ~$+$~H \citep[e.g.][]{elitzur78, neu94}. Provided that the gas density is below the critical value ($n_{\rm crit} \lesssim 10^{11}$ cm$^{-3}$) for collisional de-excitation, a sufficiently high density ($n_{\rm H_{2}} \gtrsim 10^{7}$ cm$^{-3}$) is also crucial to make maser pumping efficient.

In addition to the suitable physical conditions of the gas, the presence of cold dust in the gas cloud is also important for producing the large maser luminosities \citep{kuo18_SED} observed in water megamasers \citep{lo05}. Given that the gas temperature and density fall in the preferred ranges for population inversion, \citet{cw95} demonstrated that the maser emission can be significantly enhanced if cold dust grains exist in the cloud with a dust temperature $T_{\rm dust}=T_{\rm H_{2}}-\Delta T$, where $\Delta T \sim 50-100$ K. The presence of such cold dust can absorb nonmasing far-infrared water lines trapped in the cloud, enabling a much larger extent of H$_{2}$O molecules maintaining population inversion without being quenched \citep{goldreich1974,cw95, lo05}.

To maintain a sufficient temperature for efficient maser pumping, it has been proposed that X-rays from the active nucleus could be the primary heating source for the masing gas \citep{neu94, nm95, herrn05}. In addition, spiral shock waves travelling through a circumnuclear disk \citep{mm98} have also been proposed to pump the masers, but the velocity drifts among high-velocity features predicted by this model have not been detected in sensitive spectral-line observations of eleven H$_{2}$O maser disks \citep{dom15}, making this model less favorable \citep[see also][]{bragg2000, hum08}. Finally, recent observations of the maser emission from NGC 4258 with space VLBI \citep{baan2022} suggest that viscous heating due to the magnetorotation instability \citep[MRI; e.g.][]{arm22} could also be one of the pumping mechanisms for H$_{2}$O megamaser emission and may explain some spectral variations seen in NGC 4258. Nevertheless, including the MRI-based viscous heating in the modeling of gas disks that could produce maser emission is significantly beyond the scope of this paper and will be deferred to future work. As a consequence, we will focus in the following discussion on the scenario in which H$_{2}$O maser emission arises in the X-ray-dominated region \citep[XDR;][]{mht96,lo05} in a gas disk, within which the heating of the gas and the chemical composition are dominated by X-rays. 
%{\bf \textcolor{BrickRed}{The consideration of the possible consequences of the viscosity-based heating will be deferred to future works.} }

%and discuss the disk based on the physical conditions of the X-ray dissociation regions \citep[XDR;][]{mht96} of the disk.  

\subsection{The Role of X-ray Heating}\label{sec:3.2}

Considering a circumnuclear disk illuminated by the central X-ray source, it is expected that the disk can receive X-ray heating for maser excitation most efficiently if the disk is warped, allowing one side of the disk plane to be irradiated by X-rays directly. In the well-known picture of maser excitation in NGC 4258 \citep[][hereafter NM95]{nm95}, it was suggested that the outer edge of the warped maser disk is determined by the critical radius beyond which the disk is more directly illuminated and X-ray irradiation becomes strong enough to dissociate all molecules into atoms. By assuming the viscous gas disk to be in a steady state of accretion, NM95 show that this critical radius can be expressed as

\begin{equation}\label{eq:1}
R_{\rm cr} = 0.040L_{41}^{-0.426}(\dot{M}_{-5}/\alpha)^{0.898}\mu^{-0.383}M_{8}^{0.617}~\text{pc~, }
\end{equation}
where 10$^{41}L_{41}$ ergs s$^{-1}$ is the 2–10 keV X-ray luminosity of the central source, 10$^{-5}\dot{M}_{-5}/\alpha$ $M_{\sun}$~yr$^{-1}$ is the mass accretion rate normalized by the conventional $\alpha$ ($\lesssim 1$) viscosity parameter, and 10$^{8}M_{8}$ $M_{\sun}$ is the BH mass of the maser system. $\mu$ in the above equation is the obliquity parameter defined as $\mu$ $=$ cos~$\eta$ , where $\eta$ is the angle at which the disk is illuminated obliquely with respect to the normal direction of the disk. 

To explain the presence of the inner edge of the maser emissions in NGC 4258, NM95 observed that the maser disk appears to flatten out close to the inner radius \citep{miyoshi95}, suggesting that the obliquity parameter $\mu$ falls to zero and the disk is no longer directly illuminated by the X-ray source, making the gas too cold to mase. As a result, they speculated that the inner edge of the maser disk is determined by the nature of the warp.  While the NM95 model seems to provide a plausible explanation for the inner and outer boundaries of the maser emissions in NGC 4258 \citep[e.g.][]{martin08}, it is not yet well explored whether this model can be applied to H$_{2}$O maser disks discovered in the past two decades, whose intrinsic AGN luminosities are 2$-$3 orders of magnitude higher than that of NGC 4258.

Given the possibility that the H$_{2}$O maser disks do not necessarily follow steady-state accretion \citep[e.g.][]{gammie99}, \citet{gao17} proposed a simple alternative model in which the outer edge of a maser disk is determined by the radius beyond which $T_{\rm H_{2}} \lesssim 400$ K. Assuming the gas and dust are well-coupled in a maser disk, \citet{gao17} uses the dust temperature $T_{\rm d}$ in the optically-thin limit as a proxy for $T_{\rm H_{2}}$, finding that the gas temperature is a decreasing function of radius, with the outer radius described by $R_{\rm out} \propto L_{\rm bol}^{1/2}$, where $L_{\rm bol}$ is the bolometric AGN luminosity. 

While the derived scaling between $R_{\rm out}$ and $L_{\rm bol}$ is broadly consistent with the observations, we find that the simple approximation of $R_{\rm out}$ in \citet{gao17} can be further improved by considering the effect of X-ray heating on the gas. It is well-known that the collisions between gas and dust particles do not guarantee $T_{\rm d} \approx T_{\rm H_{2}}$ in an X-ray irradiated gas. The gas and dust can coexist at substantially different temperatures, with the difference varying depending on the the X-ray heating rate \citep[e.g.][]{dww98,lo05,nm16}. In addition, in the region where the gas density is within the favored range for maser pumping, the obscuring column density is required to be large enough \citep[e.g. $N_{\rm H}\gtrsim10^{23}$ cm$^{-2}$; ][]{kke99} to avoid molecular dissociation \citep{neu94, neu2000}. As shown in \citet{nm16}, photoheating and thermal emission from dust grains in such a regime is unimportant due to large optical depth. The dust temperature is mainly determined by collisional energy transfer between gas and dust, with $T_{\rm d}\lesssim200-300$ K \citep[see Section 2 in][]{nm16}, suggesting that using $T_{\rm d}$ in the optically-thin limit to estimate $T_{\rm H_{2}}$ may lead to non-negligible systematic errors. To obtain a more reliable estimate of $R_{\rm out}$ for a maser disk based on the physical conditions of the interstellar medium, it would be beneficial to explore the gas temperature distribution directly with the X-ray heating rate. 
%, the slope derived between Rout and mid-infrared luminosity should be 0.5, with an intercept of 22.18. Our current best-fit result is consistent with such a slope and intercept,

%%********mention Martin 2008 paper for the outer edge of NGC 4258
\begin{figure}
\begin{center} 
%\vspace*{-0.3 cm} 
\hspace*{-0.3 cm} 
\includegraphics[angle=0, scale= 0.7]{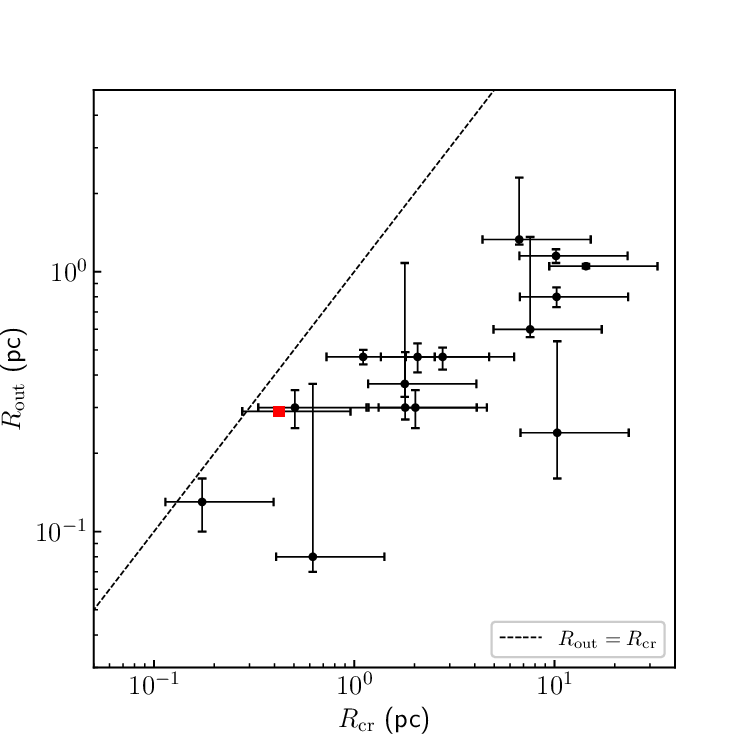} 
\vspace*{-0.5 cm} 
\caption{The comparison between the observed outer radius $R_{\rm out}$ of each maser disk and the corresponding critical radius $R_{\rm cr}$ predicted from the NM95 model. The critical radius is calculated with $\alpha=0.25$ and $\epsilon=0.42$, where $\alpha$ and $\epsilon$ are the viscosity parameter and accretion efficiency, respectively (see Section \ref{sec:3.2} \& \ref{sec:3.3}). The black dashed line indicates the locations where $R_{\rm out} = R_{\rm cr}$. Excepting three sources including NGC 4258 (the filled red square), the maser disks have $R_{\rm cr}$ substantially greater than $R_{\rm out}$. 
 }                          
\label{fig:critical_radius} 
\end{center} 
\end{figure}

\subsection{Are Disk Radii Solely Determined by X-ray Irradiation ?}\label{sec:3.3}

To examine whether the outer edge of H$_{2}$O maser emissions in the maser disks in our sample lie at the molecular-to-atomic transition radius $R_{\rm cr}$ as suggested by the NM95 model, we apply Equation \ref{eq:1} to all sixteen maser disks listed in Table \ref{tab:1} and compare the predictions with the observed values. When evaluating $R_{\rm cr}$ for our sample, we first estimate the 2-10 keV X-ray luminosity $L_{\rm X}^{2-10}=10^{41}L_{41}$ ergs s$^{-1}$ by assuming $L_{\rm X}^{2-10}=0.1L_{\rm bol}$. For all sources except for ESO 558-G009 and CGCG 074-064,  we inferred $L_{\rm bol}$ from the reddening-corrected [OIII] luminosities, $L_{\rm [OIII]}$, of the maser galaxies using bolometric corrections from \citet{heckman04} and \citet{heckman14} \citep[see][]{kuo20a}. For the two exceptions which do not have $L_{\rm [OIII]}$ available, we estimate the $L_{\rm bol}$ based on the mid-infrared AGN luminosities derived from
SED-fitting \citep{kuo18_SED}, with the bolometric correction given by \citet{gandhi09}. Since NM95 assumes a steady-state accretion disk, suggesting that the accretion rate is constant with radius, we calculate the mass accretion rate with $\dot{M}$=$L_{\rm bol}/\epsilon c^{2}$, where $\epsilon$ is the accretion efficiency for a Kerr black hole \citep[e.g.][]{cap07}, which ranges from 5.7\% to 42\% for the BH spin between $0 \le a^{*} \le 1$. To obtain the obliquity parameter $\mu$, we performed 3-dimensional Bayesian modeling for all maser disks except NGC 5495 and NGC 1068 using the MCP modeling code described in \citet{reid13}, \citet{hum13}, \citet{kuo15}, and \citet{gao16}. These modelings\footnote{In seven systems listed in Table 1, the maser acceleration measurements required for our modeling are not available. For these cases, we performed modeling by assuming the high-velocity maser features lie within $15^{\circ}$ of the mid-line of the disk plane, a typical value expected for maser disks \citep[e.g.][]{kuo11,gao16,gao17}. } are described by 15 global fitting parameters that include the galaxy distance $D$, BH mass $M_{\rm BH}$, the position of the dynamical center, and parameters characterizing the inclination warp and position angle warp \citep[see Equations (1) \& (2) in ][]{kuo13}. For all cases, the warped disk models can well fit the observational data of the maser systems with reduced $\chi_{\nu}^{2}\sim 1$ \citep[e.g.][]{kuo18_accretion}. The best-fit warp parameters enable us to calculate the obliquity parameter with $\mu=\hat{n}\cdot\hat{r}$, where $\hat{n}$ and $\hat{r}$ are the unit vectors of the disk normal and the unit radial vector for a high-velocity maser spot located at a radius $r$, respectively \citep{herrn05}. For NGC 5495 and NGC 1068, we simply make a crude estimation of $\mu$ based on their maser maps \citep[e.g.][]{gao17, greenhill97, gallimore23}, because the quality of the VLBI map for NGC 5495 is not good enough for a reliable 3-D modeling whereas NGC 1068 may have gas distribution and kinematics that deviate from the simple assumptions made in our code. 

In Columns (8), (9), \& (10) in Table \ref{tab1}, we list the obliquity parameters for the inner ($\mu_{\rm in}$) and outer ($\mu_{\rm out}$) edges of each maser disk as well as the ratio $\varepsilon_{\rm \mu}=\mu_{\rm in}$/$\mu_{\rm out}$. Based on the values of $\mu_{\rm in}$/$\mu_{\rm out}$, we see no evidence of disk flattening toward the inner edges of all maser disks (i.e. $\mu_{\rm in}$/$\mu_{\rm out}$ $\rightarrow 0$) except for UGC 3789. The ratio is of order unity for the majority of the H$_{2}$O maser disks, implying that the X-ray heating at the inner edge of the disk would not be significantly less than that at the outer edge. Therefore, under the conjecture proposed by NM95, it is difficult to account for the presence of the inner edges of most maser disks in our sample. 

In Figure \ref{fig:critical_radius}, we compare the critical radius $R_{\rm cr}$ with the observed $R_{\rm out}$ for each maser disk, with $R_{\rm cr}$ calculated based on the assumption of $\alpha=0.25$ and $\epsilon=0.42$ \citep[i.e. the maximum accretion efficiency corresponding to BH spin a$^{*}=1$; ][]{cap07}. The dashed line shows the locations where $R_{\rm out}=R_{\rm cr}$. The error bar in $R_{\rm cr}$ indicates the possible range of the critical radius given the preferred range of $\alpha\sim 0.1-0.4$ suggested from observations \citep[e.g.][]{king07, kl12}. This comparison shows that except for three sources including NGC 4258 (the red filled square in Figure \ref{fig:critical_radius}), the calculated molecular-to-atomic transition radius is substantially larger than the observed outer radius for the majority of our H$_{2}$O maser disks. %If one adopts a smaller accretion efficiency (i.e. $\epsilon<0.42$) that corresponds to a smaller BH spin (i.e. $a^{*}<1$), $R_{\rm cr}$ would become even greater than $R_{\rm out}$. 

We note that it is not unexpected for maser emission to lie well within the transition radius because the presence of molecules is of course a necessary condition for maser action. Nevertheless, the systematic difference between $R_{\rm cr}$ and $R_{\rm out}$ for the majority of the maser disks suggests that it is not the boundary of molecular dissociation in these systems, but rather some other effects that are more relevant for setting the outer radius of the maser disk. It is likely that physical conditions of the gas, such as $T_{\rm H_{2}}$ and $n_{\rm H_{2}}$, may have become unfavorable for supporting maser action before reaching the radius of molecular-to-atomic transition. Alternatively, it is also possible that some of the basic assumptions in the NM95 model, such as steady-state accretion, may not hold \citep[e.g.][]{gammie99} in some maser disks, requiring models that relax these assumptions for a better explanation of the observations. 

%, suggesting that the NM95 model could not well explain the outer radii of the majority of the maser disks. It is likely that the discrepancy between $R_{\rm cr}$ and $R_{\rm out}$ originates from the deviation from the steady-state assumption in the NM95 model. \dom{I don't follow how the steady-state assumption is determined to be the limiting factor in the NM95 model.  The values for $R_{\text{cr}}$ are systematically larger than those measured for the outermost maser spot in all systems; this seems to me like it could actually be interpreted as evidence in favor of the NM95 model?} Alternatively, it is also possible that the outer radius of the maser disk is determined by some other mechanisms, such as the the minimum temperature or density requirements for maser excitation. Both possibilities will be tested with our model presented in the following section.

\vspace{-0.2 cm}
\begin{table*}
%\small
\caption{Properties of the H$_{2}$O Megamaser Systems}\label{tab2}
    \centering
    \begin{tabular}{l c c c c c c  c c  c }
  \hline\hline       
 Name & Distance & log$L_{\rm bol}$ &  $M_{\rm D}$   &   $M_{\rm D}$  & $M_{\rm D}$ &  $\tilde{M}_{\rm D}$  &  $\tilde{M}_{\rm D}/M_{\rm BH}$   & $R_{\rm sub,Nenkova}$  & Regime    \\ 
 &  (Mpc)     &  (ergs~s$^{-1}$) &    $(s=-1.5)$    & $(s=-1.0)$       &   $(s=-0.5)$ &  ($r <= $ 1 pc)  &   ($r <= $ 1 pc)     &  (pc)   &   \\ 
\hline
   NGC4258   &  7.6  &   41.88 &    0.15  &  0.07  &  0.05  & 0.25  & 0.00006 &  0.011$\pm$0.003 &  H(D) \\ 
   NGC1068   &  13.5   &  45.58   & 24.4 &  12.2   &  8.12  & 11.6 & 0.0068 & 0.78$\pm$0.211   &    H(D)\\ 
   NGC2273   &  25.7 &   43.28  &   0.11  &  0.05  &  0.04  & 0.67  & 0.0009 &    0.055$\pm$0.015  &  H  \\
    IC2560   &  44.5 &    43.78  &  2.36  &  1.18  &  0.79  & 2.51  & 0.0057  &  0.098$\pm$0.027  &  H(D) \\
   UGC3789   &  45.4 &    44.08  &   2.20  &  1.10  &  0.73  & 3.67 & 0.0035 &   0.139$\pm$0.037  &  H \\ 
   NGC1194   &  53.2 &    43.44  &   22.9 &  11.46 &  7.64  &  8.62 & 0.0013 & 0.066$\pm$0.018  &  D \\ 
   NGC3393   &  56.2  &    44.08  &  1.84  &  0.92  &  0.61  & 1.53  & 0.0005 & 0.139$\pm$0.037   &  H/D  \\ 
J0437+2456   &  65.3  &    42.54  &   0.14  & 0.07   &  0.05  & 0.53  & 0.0018 & 0.024$\pm$0.006  &  H \\ 
   NGC2960   &  72.2 &   43.28  &   0.51  & 0.25  &  0.17  & 0.68  & 0.0006 & 0.055$\pm$0.015  &  D \\ 
   NGC5495   &  95.7 &  42.58  &   0.35  & 0.18   &  0.12  & 0.59  & 0.0006 & 0.025$\pm$0.007  &  H(D) \\ 
CGCG074-064  &   105.8 &  43.21   & 0.81 &  0.41   &  0.27  & 0.86  & 0.0004 &  0.051$\pm$0.014  & D \\
   NGC6323   &  106.0 &   43.94  &  1.14  & 0.57   &  0.38  & 1.91  & 0.0020  &  0.118$\pm$0.032  & H \\ 
ESO558-G009  &  107.6 &   43.19  &   0.97  & 0.48   &  0.32  & 1.03  & 0.0006 &  0.05$\pm$0.013  &  D \\    
  NGC5765b   &  117.0  &   44.38  &   13.5 &  6.77   &  4.51  & 5.88  & 0.0013  & 0.196$\pm$0.053 & D \\ 
   NGC6264   &  139.4 &    44.84  &   12.8 & 6.39    &  4.26  & 7.99  & 0.0027 & 0.333$\pm$0.09  &  H \\ 
   UGC6093   &  153.2 &    44.38  &   0.39  & 0.19   &  0.13  & 0.80  & 0.0003 & 0.196$\pm$0.053   & H  \\

\hline   

    \end{tabular}   
\newline
 \raggedleft 
\footnotesize{{\bf Note.} Column (1): source name; Column (2): source distance in units of Mpc; Column (3): the AGN bolometric luminosity. Except for ESO558-G009 and CGCG074-064, we adopt $L_{\rm bol}$ from \citet{kuo20a}, which are estimated from the [OIII] luminosity using the bolometric corrections from \citet{heckman04} and \citet{heckman14}. For ESO558-G009 and CGCG074-064, we estimate their $L_{\rm bol}$ based on the mid-infrared AGN luminosity derived from SED-fitting \citep{kuo18a} with bolometric correction provided by \citet{gandhi09}; Column (4)$-$(6): the mass of the disk in units of 10$^{4} M_{\odot}$ within the outer radius of the maser disk for three representative power-law indices $s=-1.5$, $s=-1.0$, and $s=-0.5$, respectively; Column (7): the disk mass in units of 10$^{4} M_{\odot}$ within 1 pc of the disk assuming the $s=-1$ model; Column (8): the mass ratio between the disk within 1 pc and the black hole; Column (9): the dust sublimation radius prescribed by \citet{nenkova08a}; Column (10): The regime into which a maser system falls. The letters H and D indicate that the outer boundaries of the maser disks are primarily limited by $R^{\rm H}_{\rm out}$ and $R^{\rm D}_{\rm out}$, respectively; H(D) indicate the cases in which $R^{\rm H}_{\rm out}$ is the primary limiting factor for the outer radius but $R^{\rm D}_{\rm out}$ is just slightly larger than $R^{\rm H}_{\rm out}$. H/D shows the special case in which $R^{\rm H}_{\rm out}$ and $R^{\rm D}_{\rm out}$ are approximately the same, implying that both heating rate and density requirements play important roles for limiting $R_{\rm out}$.    }\newline
    \label{tab:2}
\end{table*}

% ; H(D) represents the case in which the maximum heating rate is the primary limiting factor, but the outermost radius of the minimum density curve is only $\sim$10\% larger than that of the heating rate curve; H/D indicates that the outermost radius of the minimum density and maximum heating rate curves are approximately the same.

\subsection{The Outer Radius Determined by the Physical Conditions of the Gas}\label{sec:3.4}

\begin{figure} 
\begin{center} 
%\vspace*{-0.3 cm} 
\hspace*{0 cm} 
\includegraphics[angle=0, scale= 0.65]{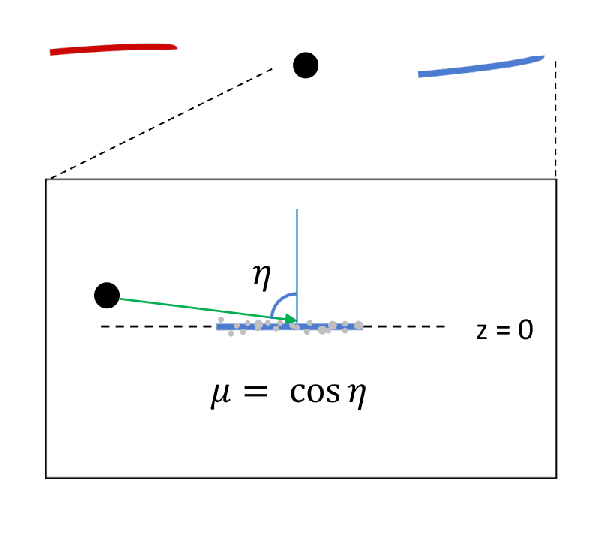} 
\vspace*{0.0 cm} 
\caption{The orientation of the warped disk in the chosen coordinate system. The plot on the top of the figure shows a representative warped maser disk, with the black circle and the red/blue lines indicating the black hole and the redshifted/blueshifted maser features in the maser disk, respectively. To evaluate the X-ray heating rate in the gas disk, we choose the coordinates such that the mid-plane of an approximately flat section of the warped disk lies in the $z=0$ plane (see the plot in the inset). The X-rays from the black hole vicinity (the green arrow) are incident on the disk at an angle $\eta$ with respect to the normal direction of the disk, where $\eta=\eta(r)$ is a function of radius $r$ due to the warped nature of the disk. In this configuration, the gas clumps (represented by the grey circles) on the $z<0$ side of the disk are expected to see more X-ray obscuring material compared to those on the $z>0$ side of the disk.
 }        
\label{fig:3}
%\label{fig: 3 } 
\end{center} 
\end{figure}

Considering a dusty, warm medium where the molecular gas is mainly heated by X-rays, instead of using $T_{\rm d}$ to approximate $T_{\rm H_{2}}$ as in \citet{gao17}, we use the X-ray heating rate to probe the gas temperature in a gas disk. This allows us to examine whether the observed outer edge of a maser disk can be explained by the minimum temperature requirement for efficient maser action as suggested by \citet{gao17}.

%It is well known that in a medium subject to X-ray heating, one can obtain the gas temperature by balancing the rates of heating and cooling \citep[e.g. emissions from CO and H$_{2}$O molecules plus gas-grain collisional cooling;][]{nm16}. \dom{Shouldn't a correct accounting of the heating+cooling balance permit you to determine the temperature any system (i.e., not just a medium subject to X-ray heating)?} 

As discussed in the beginning of the section, the water level populations for a saturated maser are primarily determined by $T_{\rm H_{2}}$, $n_{\rm H_{2}}$, $T_{\rm d}$, and $x_{\rm H_{2}O}$ in a medium optically thick to far-infrared continuum emission. If one considers an X-ray irradiated medium within which water maser emission can occur, assuming a typical value of the water abundance (e.g. $x_{\rm H_{2}O} \sim 10^{-4}$) that would arise at $T_{\rm H_{2}}\gtrsim 400$ K, it has been shown by \citet{neu2000} that one can use the equilibrium X-ray heating rate per hydrogen nucleus $H_{\rm X}/n_{\rm H}$ to replace the gas temperature $T_{\rm H_{2}}$ as one of the four key variables that determine the level populations \citep[e.g.][]{ww97, bp04}. 

In their work \citep[i.e.][]{neu2000}, it is demonstrated that the {\it maximum} heating rate that allows for efficient maser action is $(H_{\rm X}/n_{\rm H})_{\rm max}\sim 1.2\times 10^{-28}$ ergs~cm$^{3}$~s$^{-1}$, beyond which the gas will be subject to molecular dissociation. In addition, the {\it minimum} heating rate ($H_{\rm X}/n_{\rm H})_{\rm min}$ that ensures $T_{\rm H_{\rm 2}} \gtrsim 400$ K ranges from  $\sim$$5.0\times 10^{-31}$ ergs~cm$^{3}$~s$^{-1}$ to $\sim$$6.3\times 10^{-30}$ ergs~cm$^{3}$~s$^{-1}$, depending on $T_{\rm d}$. If the heating rate of a gas is between $(H_{\rm X}/n_{\rm H})_{\rm min}$ and $(H_{\rm X}/n_{\rm H})_{\rm max}$, it is expected that the gas temperature would fall within $T_{\rm H_{\rm 2}} \sim 400-1500$ K, the favored range for maser excitation. In addition, $T_{\rm H_{2}}$ would be an increasing function of $H_{\rm X}/n_{\rm H}$ if the dust temperature and water abundance are roughly constant in the region. In the following analysis, we will assume $T_{\rm d}\sim$ 300 K and $x_{\rm H_{2}O}\sim 10^{-4}$ in the masing medium, suggesting $(H_{\rm X}/n_{\rm H})_{\rm min} \sim 3.2\times 10^{-30}$ ergs~cm$^{3}$~s$^{-1}$ \citep[see Figure 1 in][]{neu2000}. For the purpose of determining the outer radius of a maser disk, we will simply identify the region within a disk where the X-ray heating rate and gas density fall within the allowed ranges that permit efficient maser action. We will not evaluate the exact gas temperature in the masing region because it will not affect our conclusion.  

Following \citet{mht96}, we compute the X-ray heating rate per hydrogen nucleus for a gas cloud in the X-ray dominated region as
\begin{equation} \label{eq:2}
H_{\rm X}/n_{\rm H}=3.8\times 10^{-25}\xi_{\rm eff} ~~\text{\rm ergs~cm$^{3}$~s$^{-1}$},
\end{equation}
where $n_{\rm H}$ is the density of the hydrogen nuclei and $\xi_{\rm eff}$ is the effective ionization parameter defined as

\begin{equation} \label{eq:3}
\xi_{\rm eff}=1.26\times 10^{-4}{F_{\rm X} \over n_{5}N_{22}^{0.9}}~.
\end{equation}

Here, $F_{\rm X}=L_{\rm X}/4\pi r^{2}$ is the {\it unattenuated} X-ray flux received by a gas clump at a radius $r$ from the central black hole with an intrinsic $1-100$ keV X-ray luminosity of $L_{\rm X}$. The two terms in the denominator show the normalized values for the total densities of hydrogen nuclei $n_{5}=n_{\rm H}/10^{5}$ cm$^{-3}$ and the X-ray attenuating column for the gas clump $N_{22}=N_{\rm H}/10^{22}$ cm$^{-2}$. 

To explore the region in the disk where the physical conditions of the gas are suitable for maser excitation, we consider the scenario in which an initially flat, cold molecular disk is warped by a certain mechanism such as {\it resonant relaxation} \citep[e.g.][]{tal09,tal12}, which can warp a sub-parsec scale disk efficiently in a timescale of $\sim$10$^{7}$ years. After the disk gets warped, one side of it will be subject to direct X-ray illumination that would deposit thermal energy in the disk. To evaluate the density distribution,  we assume that the gas motion is dominated by turbulence \citep{www98,www99}, with the turbulence velocity $c_{\rm g}\sim 2$ km~s$^{-1}$, the typical width of the water maser lines seen in Keplerian maser disks \citep[e.g.][]{reid13,kuo13,gao16}. 

To facilitate a simpler calculation, we describe the gas distribution in cylindrical polar coordinates $\vec{r}=(r,\phi,z)$ and adopt an orientation of the coordinates such that the approximately flat section of the warped disk that could produce maser emissions is aligned with the $z=0$ plane of the coordinate system (see Figure \ref{fig:3} for an illustration). Following \citet{nm95} and \citet{herrn05}, we take into account the effect of the warp on the X-ray illumination by having the central radiation incident on the disk plane at an angle $\eta$ with respect to the normal direction of the disk, with $\eta \equiv {\rm cos}^{-1}\mu$ where $\mu$ is the obliquity parameter determined by the disk warp parameters as described in Section \ref{sec:3.3}. In this geometrical setup, the BH lies slightly above the $z=0$ plane, and the effect of the variation of the warping angle as a function of radius can be accounted for by allowing the X-ray incident angle $\eta=\eta(r)$ to change with radius. Since most maser disks are slightly warped and the ratio between $\mu_{\rm in }$ and $\mu_{\rm out}$ is of order unity for the majority of our maser disks, we simply assume $\mu=\mu_{\rm out}$ in our calculation of the X-ray heating rate in the disk. The variations of $\mu$ between inner and outer radii of the maser disks are ignored in the calculation because the gradient term $d\mu/dr$ (or $d\eta/dr)$ only leads to a mild change of the heating rate distribution within the outer radius of the maser disk and does not result in any change of the conclusion of this paper.

Because of hydrostatic equilibrium, the gas density $\rho(r,z)$ at a radius $r$ from the central black hole and an elevation $z$ above the mid-plane of the slightly warped, geometrically-thin disk can be expressed as 
\begin{equation} \label{eq:4}
\rho(r,z)=\rho_{\rm mid}(r){\rm exp}\bigl[{-z^{2} \over 2H^{2}}\bigr]~,
\end{equation}
where $\rho_{\rm mid}(r)$ is the gas density at the mid-plane and $H(r)$ is the scale height of the disk, given by $H(r)=c_{\rm g}(GM_{\rm BH})^{1/2}r^{3/2}$  \citep{arm22}. The mid-plane density can be calculated with $\rho_{\rm mid}(r)=\Sigma(r)/[(2\pi)^{1/2}H]$ where $\Sigma(r)$ is the surface density of the disk. Here, we assume that the surface density takes the form of 

\begin{equation} \label{eq:5}
\Sigma(r)=\Sigma_{\rm 0}(r/a_{\rm 0})^{s}~,
\end{equation}
where the power-law index $s$ is allowed to vary in the interval of $-2<s<0$ and $\Sigma_{\rm 0}$ is the surface density at the arbitrarily chosen reference radius $a_{0}$ of the gas disk \citep{cap07,hure11,kuo18_accretion}. If $M_{\rm D}$ represents the disk mass within $a_{\rm 0}$, one finds that $\Sigma_{\rm 0}=(s+2)M_{\rm D}/2\pi a_{\rm 0}^{2}$. In the subsequent analysis, we choose the reference radius $a_{0}$ to be $a_{0}=R_{\rm out}$, and $M_{\rm D}$ thus indicates the disk mass within the outer boundary of the maser disk.

In the context of this power-law surface density profile, the steady-state accretion disk adopted by NM95 corresponds to $s=-1.5$ \citep{herrn05, kuo18_accretion}, with $M_{\rm D}$ expressed as
\begin{equation} \label{eq:6}
M_{\rm D}={ 4(GM_{\rm BH})^{1/2}\dot{M}a_{0}^{1/2} \over 3\alpha c_{\rm g}^{2} }~,
\end{equation}
where $\dot{M}$ is the mass accretion rate that can be inferred with $\dot{M}=L_{\rm bol}/\epsilon c^{2}$. Note that the disk mass $M_{\rm D}$ in the steady-state model can be fully determined by $M_{\rm BH}$, $L_{\rm bol}$, $c_{\rm g}$, $\epsilon$ and $\alpha$ whereas $M_{\rm D}$ in the power-law model described by Equation \ref{eq:5} is treated as a free parameter in our modeling. As a result, the analysis based on the power-law model would not necessarily give the same disk mass as derived from the steady-state scenario even if one adopts $s=-1.5$. If one finds that the power-law model assuming $s=-1.5$ suggests a significantly different disk mass from the value predicted by Equation \ref{eq:6}, it may imply a deviation from the steady-state accretion.

Given the above expressions, one can evaluate the number density of the molecular gas in the disk with 
\begin{equation} \label{eq:7}
n_{\rm H_{2}}(r,z)=\rho(r,z)/\zeta_{\rm H_{2}} m_{\rm H} ~, 
\end{equation}
where $\zeta_{\rm H_{2}}=2.36$ is the mean molecular weight per hydrogen molecule \citep{fd08} and $m_{\rm H}$ is the mass of the hydrogen atom.

Assuming a parcel of gas directly irradiated by the X-ray photons from the $z>0$ side of the disk is located at the position $\vec{r}=(r,\phi,z)$, we calculate the shielding column density of hydrogen nuclei in between the gas and the X-ray source as 
\begin{equation} \label{eq:8}
N_{\rm H}(r,z)=\frac{1}{\mu}\int^{\infty}_{z}n_{\rm H}(r,z')dz'\\
=\frac{\Sigma(r)}{\mu\zeta_{\rm H_{2}}m_{\rm H}}[1-{\rm Erf}(z/\sqrt{2}H)]~, 
\end{equation}
where the obliquity parameter $\mu$ accounts for the increase in the obscuring column density due to the fact that the disk is illuminated obliquely \citep[e.g.][]{nm95,herrn05}, Erf(X) is the standard error function and $n_{\rm H}(r,z)=2n_{\rm H_{2}}(r,z)$ is the number of H nuclei. For the region where Thomson scattering becomes important and enhances X-ray obscuration (i.e. the Compton-thick regime; $N_{\rm H}(r,z)$ $\ge$ 1.5$\times$10$^{24}$ cm$^{-2}$), we adopt the {\it effective} column density $N^{\rm eff}_{\rm H}(r,z)=\tau_{\rm T}N_{\rm H}$ for evaluating the attenuated X-ray heating rate, where the boosting factor is $\tau_{\rm T}\sim 6.65\times 10^{-25}N_{\rm H}$ \citep{neu2000}.  

Finally, in our calculation of $H_{\rm X}/n_{\rm H}$ with Equations \ref{eq:2} and \ref{eq:3}, we assume that the X-rays that originate from the disk corona in the vicinity of the central BH were emitted isotropically \citep{nm16} and the $1-100$ keV X-rays account for $\sim$20\% of the bolometric flux \citep[e.g.][]{neu94, netzer19}, suggesting that $F_{\rm X}=0.2L_{\rm bol}/4\pi r^{2}$. In addition to estimating the heating rate, we also compute the density distribution of the molecular gas with Equation \ref{eq:7} to find out the region where $n_{\rm H_{2}}(r,z)$ falls in the favored range for maser excitation (i.e.  $n_{\rm H_{2}}$ $=$ $10^{7}-10^{11}$ cm$^{-3}$). By comparing this region with the locations where the X-ray heating rate is sufficient to maintain $T_{\rm H_{2}}\sim 400-1500$ K, we identify the boundaries of the region within which the level population of water molecules could be inverted. The outer radius $R_{\rm out}$ of a maser disk in our model is then defined as the outermost radius of the region within which {\it both} gas temperature and density fall within the favored ranges for population inversion. We call such a region the {\it masing region}.

% Mention Masini (2016). and 2018 hydrodynamic wind paper

\begin{figure*} 
\begin{center} 
%\vspace*{-0.3 cm} 
\hspace*{-1.5 cm} 
\includegraphics[angle=0, scale= 0.7]{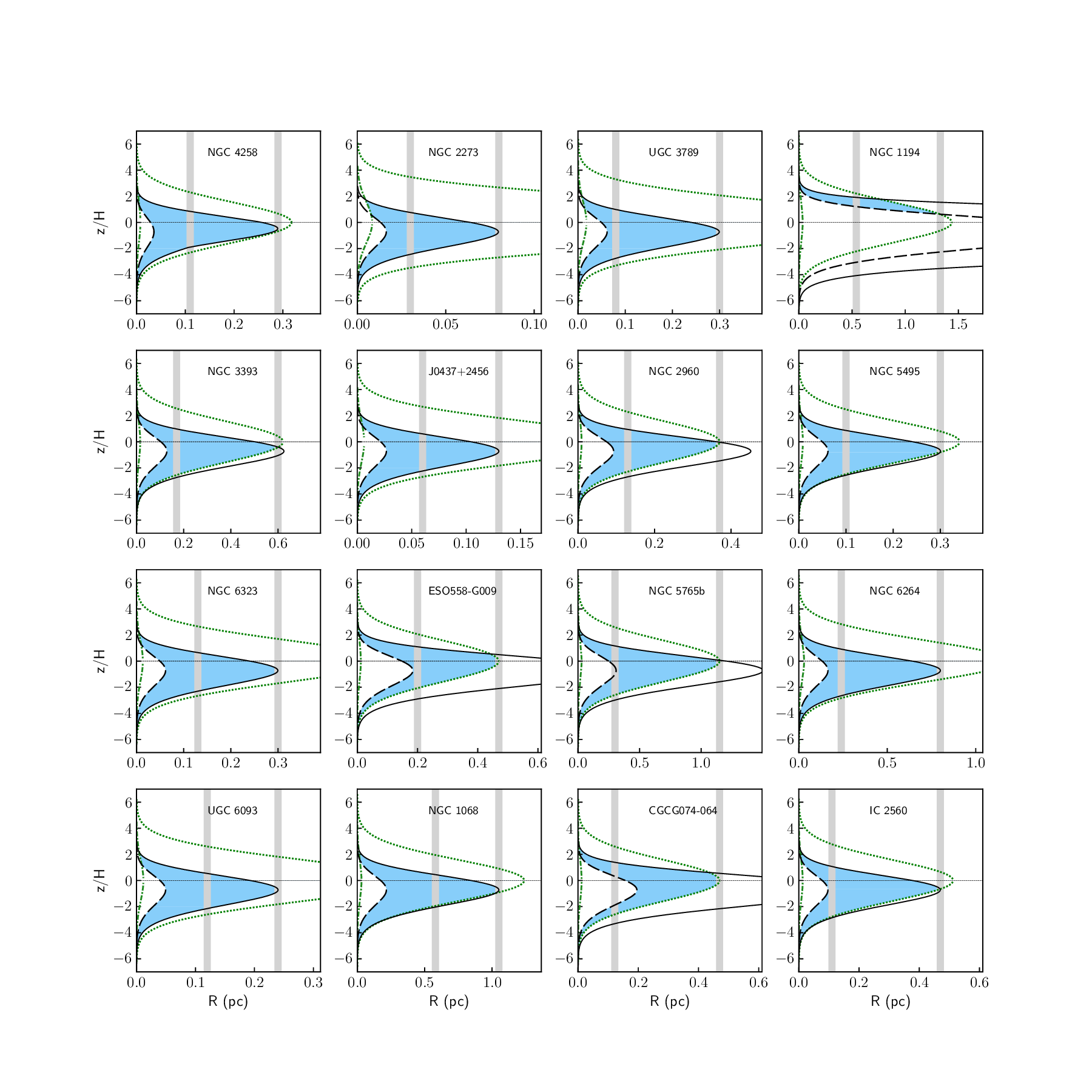} 
\vspace*{-1.0 cm} 
\caption{The distribution of X-ray heating rate per hydrogen nucleus ($H_{\rm X}/n_{\rm H}$) and the number density of molecular gas ($n_{\rm H_{2}}$) for each of the H$_{2}$O megamaser disks in our sample. The horizontal axis gives the disk radius and the vertical axis designates the disk elevation normalized to the scale height $H$. The green dotted and dot-dashed lines indicate the locations in the disk where $n_{\rm H_{2}} = 1\times 10^{7}$ cm$^{-3}$ and $1\times 10^{11}$ cm$^{-3}$, the minimum and maximum density for population inversion, respectively. The black solid line represent the points where the heating rate reaches the critical value for molecular dissociation. The dashed line shows the curve of the minimum heating rate that ensures $T_{\rm H_{\rm 2}} \gtrsim 400$ K. The blue shaded area in each plot shows the region in which both the gas density and temperature fall into the favored ranges for population inversion. The observed inner and outer radii of the maser disk are shown by the two vertical grey bars in each plot. 
 }        
\label{fig:4}
%\label{fig:4} 
\end{center} 
\end{figure*}

\begin{figure} 
\begin{center} 
\vspace*{0 cm} 
\hspace*{-0.5 cm} 
\includegraphics[angle=0, scale= 0.6]{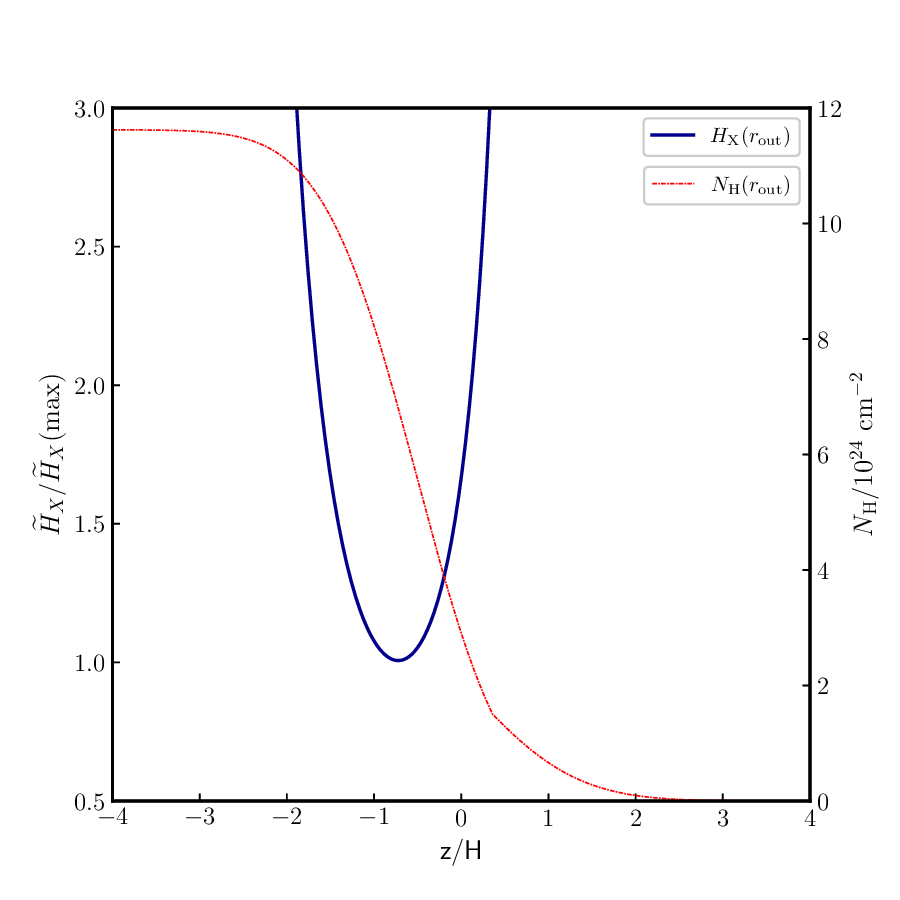} 
\vspace*{-0.5 cm} 
\caption{The representative X-ray heating rate per hydrogen ($\widetilde{H}_{\rm X}\equiv H_{\rm X}/n_{\rm H}$; the blue curve) and obscuring column density  ($N_{\rm H}$; the red dashed line) as functions of the normalized disk elevation $z/{\rm H}$ calculated with the disk parameters of NGC 2273 at its outer radius. On the y-axis, the heating rate $\widetilde{H}_{\rm X}$ is normalized by $\widetilde{H}_{\rm X}$(max), which represents the maximum heating rate that enables the presence of molecules. The column density $N_{\rm H}$ increases toward the $-z$ direction at a given radius $r$ because the $z<0$ side of the disk sees more obscuring materials than the $z>0$ side in a disk illuminated by X-rays obliquely (see Figure \ref{fig:3}). At $z\gtrsim0$, $\widetilde{H}_{\rm X}$ decreases rapidly with $-z$ as $N_{\rm H}$ rises quickly (see Equations \ref{eq:2} \& \ref{eq:3}). The heating rate reaches the minimum when the exponential drop of $n_{\rm H}$ in the $z<0$ side of the disk starts to dominate over the increase in $N_{\rm H}$. It is the interplay between $n_{\rm H}(z)$ and $N_{\rm H}(z)$ at a given radius that leads to the negative peak position and slight asymmetry of the heating rate curve.  
 }                          
\label{fig:5} 
\end{center} 
\end{figure}

\subsection{The Location and Outer Boundary of the Masing Region}

%\subsubsection{The Power-law Surface Density Model }\label{sec:3.5.2}

Given the possibility that not all maser disks necessarily follow steady-state accretion, we adopt the more general power-law disk model prescribed by Equation \ref{eq:5} for our analysis, with $M_{\rm D}$ and $s$ treated as free parameters. We model every maser disk by choosing a set of ($s$, $M_{\rm D}$), with $s$ varying between $-2.0 < s < 0.0$ in steps of 0.1. For a given value of $s$, our model suggests that the outer radius of the masing region is an increasing function of $M_{\rm D}$. By using the observed outer radius as the constraint and fixing $s$ at a specific value, we solve for the disk mass that can predict $R_{\rm out}$ consistent with the observation. 

Our analysis shows that solutions of $M_{\rm D}$ can be found for all values of $s$ between $-2.0$ and 0. In Column (4), (5), \& (6) of Table \ref{tab:2}, we list the solutions for $M_{\rm D}$ in units of $10^{4} M_{\odot}$ for three representative models with $s=-1.5$, $s=-1.0$, and $s=-0.5$, respectively, which are selected to show the typical variations of $M_{\rm D}$ when the power-law index $s$ varies by $\Delta s\sim0.5-1$. To compare the disk mass within the same reference radius, we also show in Column (7) of Table \ref{tab:2} the total disk mass $\tilde{M}_{\rm D}$ within $r=1$ pc for each maser system. In this comparison, we adopt the model of $s=-1$ \citep[i.e. the Mestel disk;][]{mestel63}, which is often adopted in the study of accretion disks \citep[e.g.][]{hure11,kuo18_accretion, gallimore23}. Given this model, the disk mass within the reference radius of $a_{0} = 1$ pc is evaluated as $\tilde{M}_{\rm D}$ $=$ $M_{\rm D}(1 ~{\rm pc}/R_{\rm out})$. 

One can see that the disk masses within 1 pc are comparable for the majority of the maser systems and the disk-to-BH-mass ratios $\tilde{M}_{\rm D}/M_{\rm BH}$ shown in Column (8) are all significantly smaller than unity, with a mean value of $\tilde{M}_{\rm D}/M_{\rm BH}\sim 0.0018$. This is consistent with the fact that most maser disks in our sample follow nearly perfect Keplerian rotation. Finally, we also note that the solutions for $M_{\rm D}$ are substantially smaller than the predictions from the steady-state accretion model, suggesting that the steady-state assumption may be in question. Given $L_{\rm bol}\sim 10^{44}$ ergs~s$^{-1}$ and $M_{\rm BH}\sim 10^{7} M_{\odot}$ for most maser disks (see Tables \ref{tab:1} and \ref{tab:2}), Equation \ref{eq:6} predicts that $\tilde{M}_{\rm D}$ would be $\sim$$1.0\times 10^{6} M_{\odot}$ if the accretion is in the steady state assuming $\epsilon=0.42$ and $\alpha=0.25$. This is $\sim$$1-2$ orders of magnitudes greater than our solutions for $M_{\rm D}$, suggesting that the steady-state accretion disk is in general too massive to yield an outer radius small enough to be consistent with those found in the observations.   

To illustrate how the physical conditions of the gas define the boundaries of the masing region, we show in Figure \ref{fig:4}  the X-ray heating rate and gas density distributions for each maser disk based on the Mestel model (i.e. $s=-1$). In each panel, the two vertical grey bars indicate the observed inner and outer radii of the maser disk. The green dotted and dot-dashed lines delineate the locations for the minimum and maximum density for maser excitation, respectively. The black solid line represents the points where the heating rate reaches $(H_{\rm X}/n_{\rm H})_{\rm max}$ while the black dashed line shows the curve for $(H_{\rm X}/n_{\rm H})_{\rm min}$. It is expected that the gas lying beyond the maximum heating curve will become atomic, and the gas bound within the minimum heating curve will be too cold to mase. To produce luminous maser emissions, the gas needs to lie within the masing region marked by the blue shaded area, in which the gas density and temperature would both fall into the favored ranges for population inversion simultaneously. Outside the masing region, either the gas temperature or density is expected to fall outside the preferred range.

To see why the maximum and minimum heating rate curves are roughly symmetric functions which peak at a negative value $z\sim -0.7$, one can solve for the analytical solutions for the heating rate curves \footnote{When deriving this equation analytically, we assume that the X-ray obscuration at $z\sim 0$ and $r \sim R_{\rm out}$ is Compton thick, which is expected to be the case for all sources except NGC 4258, according to our numerical modeling. This requires us to use the effective column density $N^{\rm eff}_{\rm H}(r,z)$ in Section \ref{sec:3.4} for our derivation. For NGC 4258, which has a very low disk mass and is Compton thin at the outer radius, skipping the Compton-thick correction $\tau_{\rm T}$ leads to a slightly different equation, but the $R_{\rm out} \propto M_{\rm BH}$ scaling remains the same. } by using Equations \ref{eq:2} through \ref{eq:8}. This allows one to express the radius of the maximum/minimum heating rate as a function of the disk elevation $z$ as
\begin{equation}\label{eq:9}
R^{\text{H}}(z) =g(H_{\rm X})^{0.44}f(\mathbb{D})e^{-0.43(z/\sqrt{2}H)^{2}}\text{Erfc}^{0.78}(z/\sqrt{2}H)~~~\text{pc}, 
\end{equation}
where Erfc($z/\sqrt{2}$H) is the complementary error function, $g(H_{\rm X})$ refers to the maximum (or minimum) heating rate per nucleus $(H_{\rm X}/n_{\rm H})_{\rm m}$ normalized by $3.19\times 10^{-28}$ ergs~cm$^{3}$~s$^{-1}$, and $f(\mathbb{D})$ is a function of the accretion disk parameters expressed as :
\begin{equation}\label{eq:10}
f(\mathbb{D}) = \Bigl(\frac{\lambda_{\rm Edd}}{0.1}\Bigr)^{-0.44}\Bigl(\frac{M_{\rm BH}}{10^{7} M_{\odot}}\Bigr)\Bigl(\frac{\tilde{M}_{\rm D}/M_{\rm BH}}{0.005}\Bigr)^{1.22}\Bigl(\frac{\mu}{0.15}\Bigr)^{-0.78}.
\end{equation}
Given a set of disk parameters, $R^{\rm H}(z)$ is dominated by the Gaussian term $e^{-0.43(z/\sqrt{2}H)^{2}}$, suggesting that it is an approximately symmetric function of $z$. By solving for the equation $dR^{\rm H}_{\rm out}/dz=0$, one obtains the peak position of the function, which occurs at $z/H=-0.76$, displaced from the $z=0$ plane by the same amount regardless of the disk parameters. The negative displacement in the disk elevation does not change with the magnitude of the disk warp because the obliquity parameter $\mu$ is independent of $z$ in the thin disk approximation assumed in our calculation. To see more intuitively the physical origin of the negative displacement and the slight asymmetry of heating rate curves, we refer the readers to the illustration in Figure \ref{fig:5}.

In Figure \ref{fig:4}, one can also see that the masing region typically lies close to the mid-plane of the disk except for NGC 1194, and the thickness of the masing region between $R_{\rm in}$ and $R_{\rm out}$ is typically $\sim$$1-4 H$. In addition, one can also infer that the X-ray heating rate per nucleus always increases with radius in every maser disk, suggesting that the gas temperature would increase with radius, while assuming $T_{\rm d}$ is roughly constant. As a result, we argue that the minimum temperature $T_{\rm min}\sim 400$ K could not be the primary factor that determines $R_{\rm out}$ as speculated in \citet{gao17}. As suggested by Figure \ref{fig:4}, the outer radius of a maser disk is primarily determined either by the maximum heating rate (e.g. NGC 4258, UGC 3789, etc.) or the minimum gas density $n_{\rm min}$ for maser pumping (e.g. NGC 2960, NGC 5765b, etc.), depending on the combination of $L_{\rm bol}$, $M_{\rm BH}$ ,$M_{\rm D}$, and the obliquity parameter $\mu$.

\subsection{The Critical Black Hole Mass}\label{sec:3.6}
%To understand why the scaling between $R_{\rm out}$ and $M_{\rm BH}$ changes in different cases, we examine the heating rate and density distributions of the maser disks associated with different $M_{\rm BH}$, $\lambda_{\rm Edd}$, and $\tilde{M}_{\rm D}$. We note that the outer boundary of the minimum density region lies beyond the maximum heating curve of the gas when $M_{\rm BH} \ll M_{\rm BH}^{\rm crit}$. On the other hand, the edge of the minimum density region is bound within the maximum heating curve if $M_{\rm BH} \gg M_{\rm BH}^{\rm crit}$. As a result, the outer radius of a maser disk is primarily determined either by the minimum density or the maximum X-ray heating, depending on whether the BH mass is greater or smaller than $M_{\rm BH}^{\rm crit}$.

To identify the most important factor that determines whether the outer radius of a maser disk is primarily limited by the maximum heating rate $(H_{\rm X}/n_{\rm H})_{\rm max}$ or the minimum gas density $n_{\rm min}$, it is helpful to look for the analytical expressions for the outermost radii of the curves for $(H_{\rm X}/n_{\rm H})_{\rm max}$ and $n_{\rm min}$ shown in Figure \ref{fig:4}. Based on Equations \ref{eq:9} and \ref{eq:10}, one can easily show that the outer radius confined by the maximum heating can be expressed as   
\begin{equation}\label{eq:11}
\begin{split}
R_{\rm out}^{\rm H}=0.81\Bigl(\frac{\lambda_{\rm Edd}}{0.1}\Bigr)^{-0.44}\Bigl(\frac{\tilde{M}_{\rm D}/M_{\rm BH}}{0.005}\Bigr)^{1.22}\Bigl(\frac{M_{\rm BH}}{10^{7} M_{\odot}}\Bigr)\\
\Bigl(\frac{\mu}{0.15}\Bigr)^{-0.78}~~{\rm pc~,}
\end{split}
\end{equation}
reminiscent of the scaling $R_{\rm out} \propto M_{\rm BH}$ found by \citet{wy12}. By combining Equations \ref{eq:4}, \ref{eq:5}, and \ref{eq:7}, one can find that the outer radius of the maser disk in the minimum density limited regime is
\begin{equation}\label{eq:12}
R_{\rm out}^{\rm D}=0.79\Bigl(\frac{\tilde{M}_{\rm D}/M_{\rm BH}}{0.005}\Bigr)^{0.4}\Bigl(\frac{M_{\rm BH}}{10^{7} M_{\odot}}\Bigr)^{0.6}~{\rm pc, }
\end{equation}
consistent with the empirical relationship $R_{\rm out} \propto M_{\rm BH}^{0.57\pm 0.16}$ found by \citet{gao17}. By equating $R_{\rm out}^{\rm H}$ with $R_{\rm out}^{\rm D}$, it can be seen that the outer radii given by the above two equations would be approximately the same ($R_{\rm out}^{\rm H} \approx R_{\rm out}^{\rm D}$) when $M_{\rm BH}\approx M_{\rm BH}^{\rm crit}$, where $M_{\rm BH}^{\rm crit}$ is a critical BH mass given by 

\begin{equation}\label{eq:13}
\frac{M_{\rm BH}^{\rm crit}}{10^{7}~M_{\odot}} \approx \Bigl(\frac{\lambda_{\rm Edd}}{0.1}\Bigr)^{1.10}\Bigl(\frac{\tilde{M}_{\rm D}/M_{\rm BH}}{0.005}\Bigr)^{-2.05}\Bigl(\frac{\mu}{0.15}\Bigr)^{1.95}.
\end{equation}

This critical BH mass separates the BH mass of the maser disk into two regimes. In the regime where $M_{\rm BH} << M_{\rm BH}^{\rm crit}$, the outer radius of the disk is primarily limited by the maximum heating rate, with the outer radius determined by $R_{\rm out}^{\rm H}$. When the BH mass of the system is $M_{\rm BH} >> M_{\rm BH}^{\rm crit}$, the maser disk falls into the minimum density limited regime where the outer radius is determined by $R_{\rm out}^{\rm D}$. So, given a set of disk parameters, the critical BH mass plays a decisive role in determining whether the outer boundary of the maser disk is mainly confined by $(H_{\rm X}/n_{\rm H})_{\rm max}$ or $n_{\rm min}$. To identify the regime for each maser system, we label our sample sources with either "H" or "D" in Column (10) in Table \ref{tab:2} to indicate whether they fall into the heating rate limited regime (H) or minimum density limited regime (D). We note that for some heating rate limited sources, their $R_{\rm out}^{\rm D}$ are comparable to $R_{\rm out}^{\rm H}$ ($R_{\rm out}^{\rm D} \gtrsim R_{\rm out}^{\rm H}$), suggesting that their $M_{\rm BH}$ are close to the critical values. We label these sources with 'H(D)' to indicate that $R_{\rm out}^{\rm D}$ is almost as limiting as $R_{\rm out}^{\rm H}$, and they could appear to follow both scaling relations mentioned above because they lie at the transition region between the two regimes. For the special case (NGC 3393) in which 
$R_{\rm out}^{\rm D}$ is nearly the same as $R_{\rm out}^{\rm H}$, we label it with 'H/D' to show that both limiting factors play equally important roles for determining $R_{\rm out}$.

The existence of the two regimes separated by $M_{\rm BH}^{\rm crit}$ gives an interesting implication that the seemingly disparate results from \citet{wy12} and from \citet{gao17} can be unified once one understands that these two regimes of BH mass exist. Indeed, when we examine the eight maser disks studied in \citet{wy12}, except for NGC 4388 which is not included in our modeling, we find that 5 of the 7 remaining sources fall in the heating rate limited regime (i.e. NGC 4258, NGC 2273, UGC 3789, NGC 6323, and NGC 6264), explaining the $R_{\rm out} \propto M_{\rm BH}$ scaling revealed in their work. On the other hand, since \citet{gao17} include several additional sources in the density limited regime (e.g. ESO 558-G009, NGC 5765b, CGCG 074-064, and NGC 1194) as well as sources for which $M_{\rm BH}\approx M_{\rm BH}^{\rm crit}$ (e.g. IC 2560 and NGC 5495), it is understandable that \citet{gao17} would reach a different scaling between $R_{\rm out}$ and $M_{\rm BH}$.

Finally, the readers should be aware that the outer radius of a maser disk predicted by our model should be seen as an approximation. In our simple modeling presented above, we ignore the effect of density perturbation of the molecular gas as a result of X-ray heating. Given the turbulence-dominated disk in our analysis, we implicitly assume that the equilibrium density of the molecular gas after X-ray irradiation is comparable to the gas density before any injection of thermal energy. Taking the density perturbation into account would require a more rigorous modeling that involves solving the energy balance equation for the gas. This is beyond the scope of this paper, and we defer this to future work.

\begin{figure} 
\begin{center} 
%\vspace*{-0. cm} 
\hspace*{-0.3 cm} 
\includegraphics[angle=0, scale= 0.43]{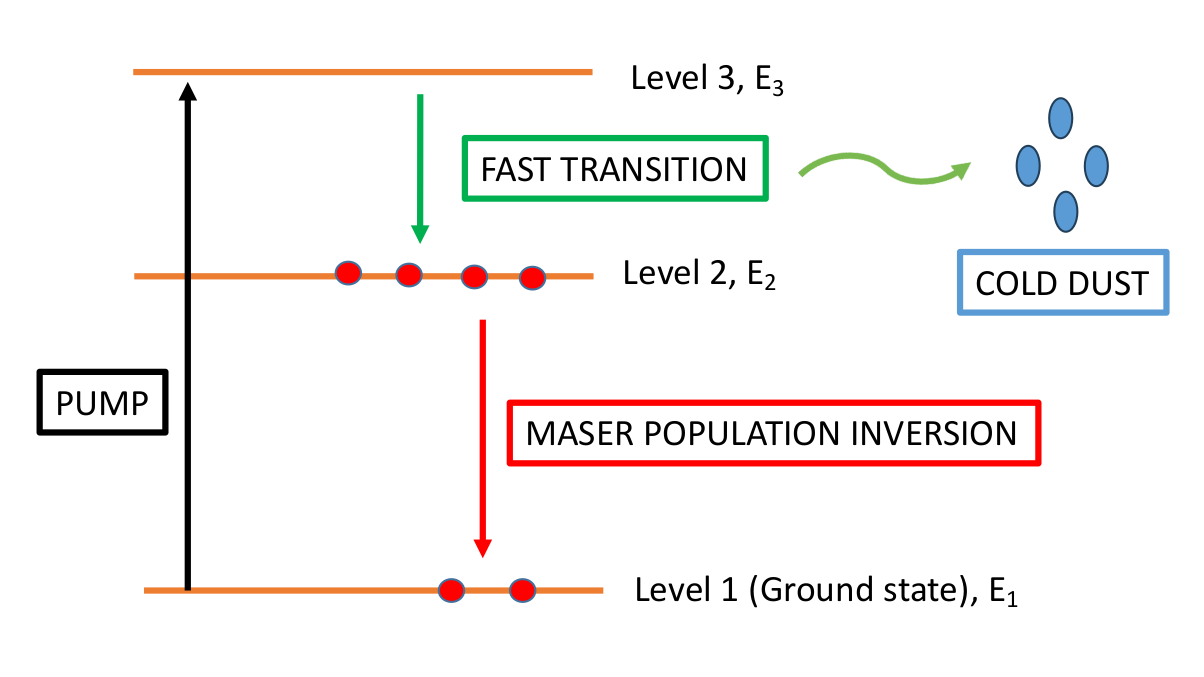} 
\vspace*{-0.3 cm} 
\caption{The schematic diagram of the maser excitation process. For the maser action to take place, the H$_{2}$O molecules are first pumped to a short-lived state at Level 3, followed by spontaneous transitions into a long-lived state at Level 2. The $\text{E}_{3}-\text{E}_{2}$ transition that releases far-infrared photons needs to be fast enough to maintain the maser population inversion between $\text{E}_{2}$ and $\text{E}_{1}$. To avoid having the far-infrared photons re-absorbed from Level 2 to Level 3, making the population moving into Level 2 effectively slow, it is useful if the emitted $\text{E}_{3}-\text{E}_{2}$ photons are removed from the maser pump cycle e.g., through absorption by cold dust.  If the dust has been eliminated by sublimation, this possibility is removed and the maser population inversion can be affected or even destroyed due to the thermalization of the population \citep[e.g.][]{lo05}.
}                       
\label{fig:6} 
\end{center} 
\end{figure}

\begin{figure} 
\begin{center} 
%\vspace*{-0.3 cm} 
\hspace*{-0.3 cm} 
\includegraphics[angle=0, scale= 0.6]{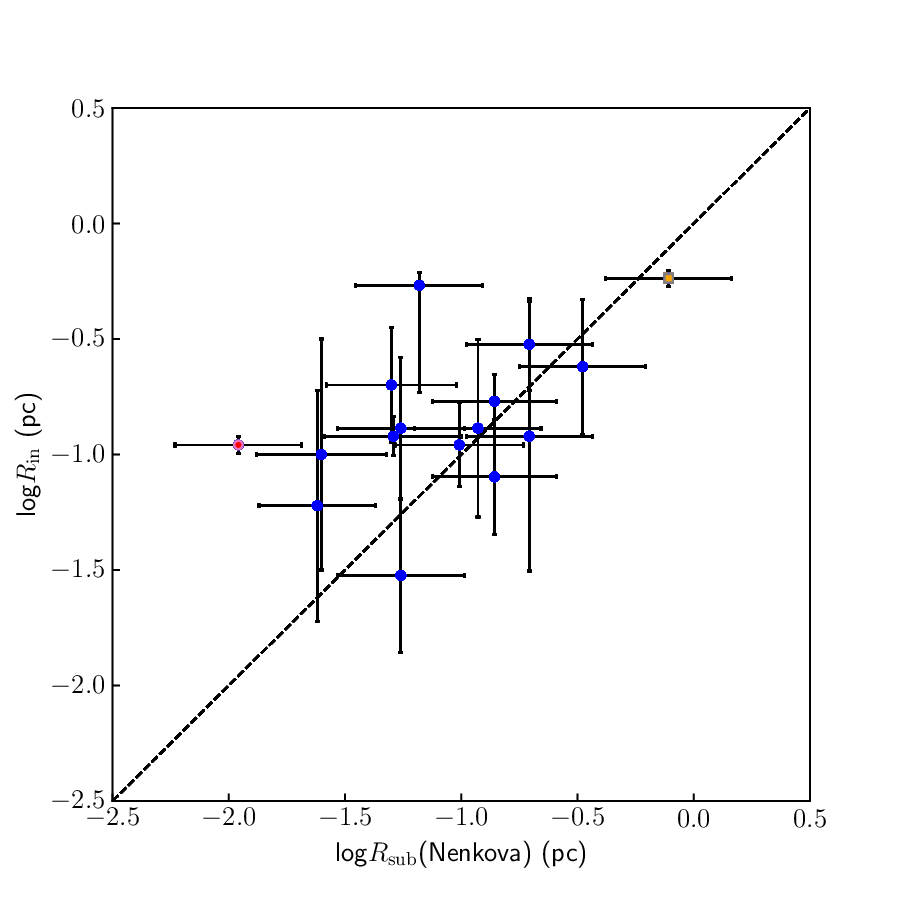} 
\vspace*{0.0 cm} 
\caption{A comparison between the observed inner radius $R_{\rm in}$ and the dust sublimation radius $R_{\rm sub}$ of the H$_{2}$O maser disks. The dashed line in the plot indicates where $R_{\rm in}=R_{\rm sub}$. Except for NGC 4258, which is marked by the filled red circle, the $R_{\rm in}$ values are comparable to $R_{\rm sub}$ for our maser disks, given the measurement uncertainties. The yellow square denotes NGC 1068, whose $R_{\rm in}$ appears to be an outlier in the right panel of Figure 1, but its inner radius is well consistent with the dust sublimation radius as shown in this figure.
 }                          
\label{fig:7} 
\end{center} 
\end{figure}

\subsection{The Inner Edge and the Dust Sublimation Radius}\label{sec:3.7}

One can see in Figure \ref{fig:4} that the density and heating rate requirements for efficient maser pumping do not impose a clear inner bound for the maser emissions in the maser disk. This situation does not change no matter how we vary the model parameters, suggesting that the physical conditions of the gas alone may not be able to define the inner boundary of a maser disk. Other factors might be involved.

Among the additional factors that could affect the production of water maser emissions, dust properties are the most important. As explained in Section \ref{sec:3.1}, the presence of cold dust is essential for maintaining population inversion by absorbing nonmasing far-infrared water lines trapped in the cloud. If the amount of dust is substantially reduced at some radius, for example because of dust sublimation, it is possible that the trapping of the far-infrared photons from non-masing water transitions may become more significant (see Figure \ref{fig:6} for an illustration). This process could substantially diminish or even destroy the population inversion, leading to a boundary of the maser emissions in the disk \citep[e.g.][]{greenhill96, kke99, galllimore2001}.

To explore this possibility, we estimate the dust sublimation radius for each maser disk (see Column (9) of Table \ref{tab:2}) based on the prescription provided by \citet{nenkova08a}:
\begin{equation}\label{eq:14}
R_{\rm sub, Nenkova} \approx 0.4\bigl({ L_{\rm bol} \over 10^{45} ~{\rm erg}^{-1} }\bigr)^{1/2}\bigl({ 1500 K \over T_{\rm sub} }\bigr)^{2.6} ~~\text{pc},
\end{equation}  
which is derived for the clumpy torus models assuming the grain mix has standard
interstellar properties \citep{nenkova2008b}. In our calculation, we use $L_{\rm bol}$ from Table \ref{tab:2}, with the assumption that dust sublimates at the temperature $T_{\rm sub} \sim 1500$ K. We estimate the uncertainty $\delta R_{\rm sub, Nenkova}$ by adopting $\delta L_{\rm bol}\sim 0.54$ dex, obtained from the comparison between $L_{\rm bol}$ estimated from X-ray spectroscopy and the [OIII]$\lambda$5007 line for our sample \citep{kuo20a}. As noted in  \citet{nenkova08a}, $R_{\rm sub, Nenkova}$ is not a sharp boundary within which no dust exists. Instead, it is an approximation that marks the radius across which the environment transitions from being dusty to dust-free as the individual components of the dust mixture gradually sublimate at different radii. It is expected that the largest grains can survive down to the innermost radius of the dusty torus probed by reverberation mapping observations \citep[e.g.][]{suga06, kishimoto07}, which is $\sim$3 times smaller than $R_{\rm sub, Nenkova}$.

In Figure \ref{fig:7}, we compare the observed inner radius of the maser disk with the dust sublimation radius for all sources in our sample. This figure shows that the majority of the maser disks have their inner radii consistent with $R_{\rm sub, Nenkova}$ within $\sim1\sigma$ ($\sim$75\% of the sources) or $\sim 2\sigma$ ($\sim$19\% of the sources) error. The only significant outlier in this comparison is NGC 4258 (the red square in the figure), for which $R_{\rm in}$ is considerably greater than $R_{\rm sub, Nenkova}$ given the measurement uncertainties. This result suggests that, except for NGC 4258 (see Section 4.2 for further discussion), dust sublimation may play a role in determining the inner radius of a maser disk.  It is likely that dust residing near the sublimation radius would become warmer and its efficiency as a heat sink \citep{bp04} for absorbing far-infrared photons from nonmasing water transitions would get reduced. As a result, the efficiency of maser amplification could get significantly reduced as the dust temperature gradually approaches the sublimation temperature at $r \sim R_{\rm  sub, Nenkova}$ \citep[see Figure 1 in ][]{gray22}, making the maser emissions substantially weaker.

In addition to the effect of dust temperature, the reduction of the dust mass could also affect the maser intensity significantly. As dust gradually sublimates away at $R\sim R_{\rm sub, Nenkova}$, maintaining the population inversion becomes more difficult because the trapping of non-masing far-infrared water line emissions by the masing clouds becomes more significant. As shown in \citet{gray22}, a halving of the dust mass fraction can lead to an extraordinary reduction in the efficiency of maser amplification, suggesting that even before the dust temperature is high enough for dust to totally sublimate away, maser emissions could have dimmed below the detection threshold. This supports the conjecture that dust properties near the dust sublimation radius may become unfavorable for efficient maser action, making dust the culprit for setting the observed inner boundary of a maser disk.

%\begin{table}
%	\centering
%	\caption{This is an example table. Captions appear above each table.
%	Remember to define the quantities, symbols and units used.}
%	\label{tab:example_table}
%	\begin{tabular}{lccr} % four columns, alignment for each
%		\hline
%		A & B & C & D\\
%		\hline
%		1 & 2 & 3 & 4\\
%		2 & 4 & 6 & 8\\
%		3 & 5 & 7 & 9\\
%		\hline
%	\end{tabular}
%\end{table}

\section{Discussion} \label{sec:4}
\subsection{The Primary Physical Factors that Affect the Characteristic Disk Radii } \label{sec:4.1}
The results of our calculations in the last section suggest that the outer radius of a maser disk could be determined either by the minimum gas density or the maximum heating rate that enables efficient maser action. Moreover, the inner edge of the maser emissions in the disk may result from the quenching of population inversion near the dust sublimation radius. While the physical conditions of the gas and the dust appear to play an important role in defining the inner and outer boundaries of a maser disk, these conditions do not seem to directly explain why the inner and outer radii of the majority of the maser disks are around $R_{\rm in} \sim 1.0\times 10^{5} R_{\rm S}$ and $R_{\rm out} \sim 3.0\times 10^{5} R_{\rm S}$ as shown in Section 2. 

To explain the characteristic inner radius of $R_{\rm in}\sim 1\times 10^{5} R_{\rm S}$ in light of the dust sublimation radius, it is helpful to replace $L_{\rm bol}$ in Equation \ref{eq:14} with
$L_{\rm bol}=\lambda_{\rm Edd}L_{\rm Edd}$, where $\lambda_{\rm Edd}$ is the Eddington ratio and $L_{\rm Edd}=1.26\times 10^{38} (M_{\rm BH}/M_{\odot}$) is the Eddington luminosity. This replacement allows one to express the dust sublimation radius in units of $R_{\rm S}$ as  

\begin{equation}\label{eq:15}
R_{\rm sub, Nenkova} = 1.05\times 10^{5}~\bigl({\lambda_{\rm Edd} \over 0.05}\bigr)^{1/2}\bigl(\frac{M_{\rm BH}}{10^{7}M_{\odot}}\bigr)^{-1/2}~R_{\rm S}~.
\end{equation}

The above equation directly implies that the dust sublimation radius in AGN in general does not equal $\sim$$1.0\times 10^{5} R_{\rm S}$ since its value depends substantially upon $\lambda_{\rm Edd}$ and $M_{\rm BH}$. However, it is well known that the BH masses for the Keplerian megamaser disks are typically $\sim$$10^{7}M_{\odot}$, likely resulting from the fact that disk megamasers are preferentially detected in local Seyfert 2 galaxies \citep{kuo11} whose BH mass function peaks at $M_{\rm BH} \approx 3\times10^{7} M_{\odot}$ \citep{heckman04}. Moreover, the Eddington ratios of the majority of the Keplerian disk maser systems fall in the narrow range of $\lambda_{\rm Edd}\sim 0.01-0.1$, with the median value $\approx 0.04$ \citep[see Table 5 and Figure 6 in][]{kuo20a}. Given that disk maser systems typically have $M_{\rm BH} \approx 10^{7} M_{\odot}$ and $\lambda_{\rm Edd} \approx 0.04$, one can infer from Equation \ref{eq:15} that the dust sublimation radii of most Keplerian H$_{2}$O maser disks would be $\sim$$1.0\times 10^{5} R_{\rm S}$, the characteristic inner radius we see in Figure \ref{fig:1}. 

To explain the characteristic outer radius, one can follow a similar approach to express the outer radii given by Equations \ref{eq:11} \& \ref{eq:12} in units of $R_{\rm S}$. For the heating rate limited regime, this gives 
\begin{equation}\label{eq:16}
R^{\rm H}_{\rm out} = 1.85\times 10^{5}~\bigl({\lambda_{\rm Edd} \over 0.05}\bigr)^{-0.44}\bigl(\frac{\tilde{M}_{\rm D}/M_{\rm BH}}{0.0011}\bigr)^{1.22}\Bigl(\frac{\mu}{0.15}\Bigr)^{-0.78}~R_{\rm S}.
\end{equation}
For the minimum density limited regime, one obtains
\begin{equation}\label{eq:17}
R^{\rm D}_{\rm out} = 2.78\times 10^{5}~\bigl(\frac{\tilde{M}_{\rm D}/M_{\rm BH}}{0.0011}\bigr)^{0.4}\bigl(\frac{M_{\rm BH}}{10^{7}M_{\odot}}\bigr)^{-0.4}~R_{\rm S}~,
\end{equation}
where the normalization factor 0.0011 is the median value of $\tilde{M}_{\rm D}/M_{\rm BH}$ for all sources. These equations suggest that the outer radii of the maser disks would be $\sim$3$\times$10$^{5}$ $R_{\rm S}$, given the typical values of $\lambda_{\rm Edd}$, $\tilde{M}_{\rm D}/M_{\rm BH}$, $M_{\rm BH}$, and $\eta$ for the majority of the Keplerian maser disks. Although the dependence of $R_{\rm out}$ on the disk parameters is more complicated, given that these four parameters all fall within relatively narrow ranges for the majority of our maser sources, it is not unexpected that the outer radii of the maser disks have similar values when expressed in units of $10^{5} R_{\rm S}$.

It is likely that the characteristic size of maser disks is deeply connected with the fine-tuning nature of the disk megamaser phenomenon and it reflects the physical state of a gas disk in a certain phase of AGN evolution, with BH masses following the population for low redshift Seyfert 2 galaxies. As suggested in \citet{anca12}, the disk megamaser phenomena may only occur in a certain (short) phase in the galaxy-AGN coevolution during which the configuration of the gas accretion transitions from an optically thin, geometrically thick accretion flow to an optically thick, geometrically thin accretion disk, which typically have Eddington ratios $\lambda_{\rm Edd} \gtrsim 0.01$ \citep[e.g.][]{gp19}. Given this speculation, the general Eddington ratio distributions for local Seyfert galaxies \citep[e.g.][]{jones16} would imply that the disk megamasers living in geometrically thin accretion disks with $\lambda_{\rm Edd}\sim 0.01-0.1$ would outpopulate the ones with $\lambda_{\rm Edd} \gtrsim 0.1$, explaining the typical range of $\lambda_{\rm Edd}$ for most Keplerian maser systems.

It can be expected that if H$_{2}$O maser disks also exist at the centers of high redshift AGNs (e.g. quasars at $z>2$) or local Seyfert galaxies with high Eddington ratios, the combination of their Eddington ratio and BH mass distribution would be considerably different (e.g. $\lambda_{\rm Edd} \gtrsim 0.1$ and $M_{\rmn BH} \gtrsim 10^{9} M_{\rm BH}$ for high-z quasars), leading to distinctly different characteristic inner and outer radii for these sources. This could explain why NGC 1068 has the largest inner radius (i.e. $R_{\rm in}= 3.57\times 10^{5} R_{\rm S}$) among maser disks in our sample, which is $\sim$4 times greater than the median value for all sources (i.e. $R_{\rm in}= 0.87\times 10^{5} R_{\rm S}$; see Figure \ref{fig:1}). Given $M_{\rm BH}=1.7\times 10^{7} M_{\odot}$ (see Table \ref{tab:1}), the Eddington luminosity of NGC 1068 is $L_{\rm Edd}=2.1\times 10^{45}$ ergs~s$^{-1}$. The corresponding Eddington ratios are 1.8 and 0.3 if its AGN bolomotric luminosities are inferred from [OIII] luminosity and absorption-corrected X-ray luminosities, respectively \citep{kuo20a}. Because of the substantially higher Eddington ratio in comparison with the Keplerian maser disks in our sample, NGC 1068 would have an inner radius substantially greater than the mean inner radius of most maser disks according to Equation \ref{eq:15}. Nevertheless, similar to most other maser disks, NGC 1068 has an $R_{\rm in}$ well consistent with $R_{\rm sub, Nenkova}$, supporting the conjecture that its inner boundary of maser emissions is constrained by the dust sublimation process. 

%{\bf To explain the outer radius of NGC 1068, we note that this maser system also has a disk-to-BH-mass ratio ($\tilde{M}_{\rm D}/M_{\rm BH} = 0.0166$) significantly higher than the median value (i.e. $\tilde{M}_{\rm D}/M_{\rm BH}=0.0013$) for the rest of the maser disks. Since the outer radius of NGC 1068 is primarily limited by the maximum heating rate, according to our model, Equation \ref{eq:16} suggests that its outer radius in units of $R_{\rm S}$ is primarily determined by $\lambda_{\rm Edd}$, $\tilde{M}_{\rm D}/M_{\rm BH}$ and $\mu$. Given that $R^{\rm H}_{\rm out}$ is most sensitive to the change in $\tilde{M}_{\rm D}/M_{\rm BH}$, the significantly higher $\tilde{M}_{\rm D}/M_{\rm BH}$ for NGC 1068 naturally leads to a substantially larger outer radius ($R_{\rm out}= 13.7\times10^{5} R_{\rm S}$) in comparison with other sources ($R_{\rm out} \sim 3\times10^{5} R_{\rm S}$). By considering the physical causes that determine the disk mass, we speculate that the larger values of $R_{\rm in}(10^{5} R_{\rm S})$ and $R_{\rm out}(10^{5} R_{\rm S})$ for NGC 1068 could both be attributed to the same physical origin (i.e. the high Eddington ratio) because a higher value of $\lambda_{\rm Edd}$ implies a higher mass accretion rate with respect to a given BH mass, which would lead to a relatively more massive disk and a higher disk-to-BH-mass ratio, unless the mass loss rate due to factors such as AGN winds becomes significant. }

\subsection{The Inner Radius of NGC 4258}

As shown in Section \ref{sec:3.7} and Figure \ref{fig:7}, NGC 4258 is the most prominent outlier ($R_{\rm in}$ $>>$ $R_{\rm sub, Nenkova}$) in our comparison between $R_{\rm in}$ and $R_{\rm sub, Nenkova}$ for maser disks in our sample. While one can infer from Equation \ref{eq:15} that the substantially smaller dust sublimation radius in NGC 4258 ($R_{\rm sub, Nenkova} = 0.011$ pc) is caused by the combination of the extremely low Eddington ratio \citep[i.e. $\lambda_{\rm Edd}\sim 0.0001$;][]{kuo20a} plus the slightly higher BH mass \citep[i.e. $M_{\rm BH}=4.0\times 10^{7} M_{\odot}$;][]{hum13}, it is unclear why the inner boundary of the maser disk cannot reach down to $R_{\rm sub, Nenkova}$ based on what we learn in Section \ref{sec:3}. This discrepancy hints that the inner edge of this well-known maser source depends on factors other than the physical conditions of the gas and dust in the disk. 

We note that if the conditions that enable the population inversion are not the determining factors that define the inner boundary for NGC 4258, it is possible that its inner radius is characterized by the lower limit in the velocity coherence length that allows for observable maser emissions. By comparing the orientations of all maser disks in our sample, it can be seen that while most maser disks are within $\sim 1^{\circ}-2^{\circ}$ from being edge-on \citep[e.g.][]{kuo11,gao16,dom20a}, NGC 4258 displays the most significant inclination warp. Based on the disk modeling by \citet{hum13}, the disk inclination at a radius $r \le R_{\rm in}=0.11$ pc is $\le$79.2$^{\circ}$, suggesting that the deviation from the edge-on configuration would be greater than 10$^{\circ}$ if there are masers residing inside the observed $R_{\rm in}$. Given such large inclinations, it is possible that the effective coherent path length would be reduced substantially, leading to weak maser emissions below the detection limit.

In addition to the effect of disk inclination, we also note that the coherence length $L_{\rm c}$ reduces linearly with the disk radius $R$ for a maser disk \citep[i.e. $L_{\rm c} \propto R$;][]{lo05}. Since the maser flux density $S_{\rm \nu}$ scales with $L_{\rm c}$ as $S_{\rm \nu} \propto L_{\rm c}^{3}$, suggesting $S_{\rm \nu} \propto R^{3}$ (see Section \ref{sec:4.3} for a detailed discussion), it can be expected that the maser flux density would drop quickly when the disk radius is smaller than a certain value at which the physical conditions of the gas no longer support the maximum maser pumping rate \citep[e.g.][]{gray16}. As a result, in a maser system with a low Eddington ratio (e.g. $\lambda_{\rm Edd} << 0.01$), the inner radius of the maser disk may not necessarily lie close to the dust sublimation radius because the maser coherence path length at $r \gtrsim R_{\rm sub, Nenkova}$ may have become too small to produce observable maser emissions before reaching down to the dust sublimation radius. For NGC 4258, it is likely that both the relatively large disk inclination and the reduced coherence length at a smaller radius play a substantial role in causing the discrepancy between $R_{\rm in}$ and $R_{\rm sub, Nenkova}$. To confirm this hypothesis requires future work with more rigorous modeling of the maser production rate as a function of disk radius based on the physical conditions of NGC 4258.

\begin{figure*} 
\begin{center} 
%\vspace*{-0.3 cm} 
\hspace*{0.0 cm} 
\includegraphics[angle=0, scale= 0.7]{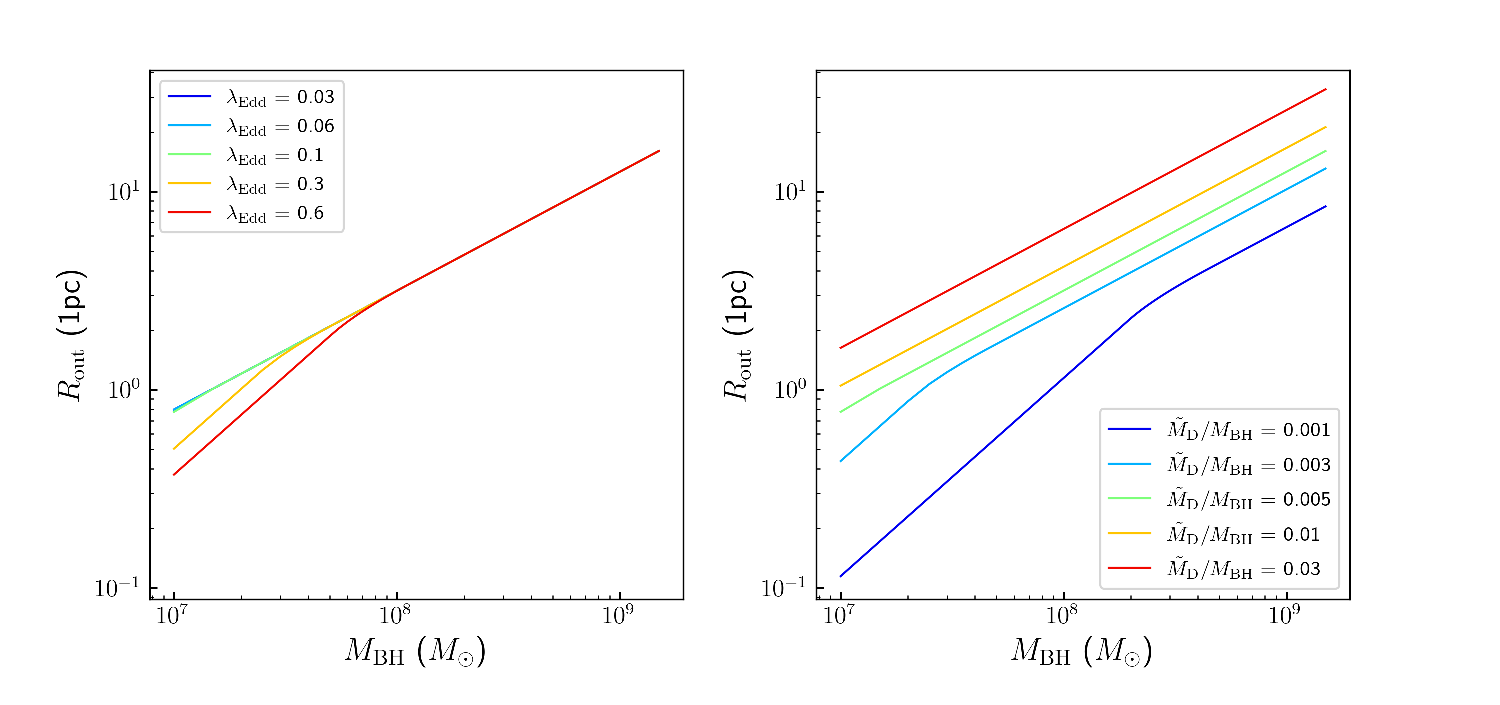} 
\vspace*{-0.3 cm} 
\caption{ Left panel: the prediction of the outer radius of a megamaser disk with the BH mass ranging from $1.0\times 10^{7} M_{\odot}$ to $1.5\times 10^{9} M_{\odot}$, made with the assumption of $\tilde{M}_{\rm D}/M_{\rm BH} = 0.005$, where $\tilde{M}_{\rm D}$ is the disk mass within one pc.  The blue, cyan, green, yellow, and red lines represent the predictions assuming Eddington ratios of 0.03, 0.06, 0.1, 0.3, and 0.6 for the accreting BH, respectively; Right panel: The outer radius of a megamaser disk given an Eddington ratio of 0.1. The blue, cyan, green, yellow, and red lines indicate the predictions calculated based on $\tilde{M}_{\rm D}/M_{\rm BH} = 0.001, 0.003, 0.005, 0.01, 0.03$, respectively. 
 }                          
\label{fig:8} 
\end{center} 
\end{figure*}

\subsection{On the Detection of H$_{2}$O Gigamasers at High Redshift}\label{sec:4.3}
\subsubsection{The Effect of Increasing Gain Length}\label{sec:4.3.1}

Observations of the early universe have revealed strong evidence that galaxies once went through an extremely rapid and luminous phase of evolution, leading to intensive star formation and quasar activity that peak at $z\sim 2-3$ \citep[e.g.][]{kot09, lapi17}. During the most active phase of evolution, it is believed that a high fraction of luminous quasars may harbor $\gtrsim 10^{9} M_{\odot}$ supermassive BHs at their cores \citep[e.g.][]{vo09, liq21}. \citet{lo05} speculated that H$_{2}$O {\it gigamasers} could be triggered in high-z active galaxies as a result of the immense energy injection into the surrounding gas, and the total maser luminosities of these gigamasers could be $\gtrsim 10^{3}$ higher than those of low-z H$_{2}$O megamasers \citep[see also][]{nm95, Koekemoer1995, Falcke2000, Impellizzeri2008, bosch16,yang2023}. If such gigamasers truly exist in the early universe, they could serve as a new class of high-redshift distance indicators, providing an independent probe of the expansion rate of the early universe. 

Over the past two decades, there have been a few maser surveys aiming to detect masers from high-z galaxies \citep[e.g.][]{Impellizzeri2008,Mckean11}. One of the surveys \citep{Impellizzeri2008} detected the first 22 GHz H$_{2}$O gigamaser system (MG J0414$+$0534) at $z>1$, confirming the existence of H$_{2}$O gigamasers in the high-z universe. However, the analysis of the maser spectra for this source \citep{Castangia11} implies that the maser emissions are more likely associated with jet-cloud interaction rather than originating from a nearly edge-on disk, which is the configuration required for precise BH mass and distance measurement. To use the maser technique for studying cosmology, one still needs to look for luminous H$_{2}$O gigamaser systems {\it in disk configurations} at high redshifts.   

To assess the possibility of discovering high-z H$_{2}$O gigamaser disks quantitatively, we estimate the flux densities of high-z maser disks based on the model presented in the last section. Our estimation suggests that existing radio telescopes, such as the Very Large Array (VLA) and the Green Bank Telescope (GBT), would have sufficient sensitivities to detect these gigamaser disks if their typical radii were $\gtrsim 20-30$ times greater than those of the low redshift maser systems (i.e. $R\sim 0.3-0.8$ pc). 

As one can infer from \citet{lo05}, the flux density $S_{\nu}$ of a saturated maser source can be expressed as  
\begin{equation} \label{eq:maser_flux}
S_{\nu} = (1+z){n_{\rm u}\Delta Ph\nu L_{\rm g}^{3} \over D_{\rm L}^{2}\Delta\nu}~ ,
\end{equation}
where $n_{\rm u}$ is the density of H$_{2}$O molecules in the upper excited state, $\Delta P$ is the rate of maser transition from the upper to the lower state, $h\nu$ is the energy of a maser photon, $L_{\rm g}$ is the gain length for maser amplification, $D_{\rm L}$ is the luminosity distance to the source, and $\Delta\nu$ is the observing bandwidth at the observer's frame. The cubic dependence on $L_{\rm g}$ shown in the equation suggests that the maser flux density is highly sensitive to the gain length, which is expected to be $L_{\rm g} \lesssim L_{\rm c}$, where $L_{\rm c}$ is the velocity coherent path length in a maser disk. In the following discussion, we will make the simple assumption that $L_{\rm g} \approx L_{\rm c}$ when trying to provide an optimistic estimation of the detectability of high-z maser disks. The readers should be aware of the caveat that $L_{\rm g}$ could be substantially smaller than $L_{\rm c}$ in cases where the gas disks are highly clumpy. In such disks, the gain length could be limited to the size of each individual cloud, and $L_{\rm g}$ would become long enough to produce luminous maser emissions detectable at high redshifts only when multiple clumpy masing clouds happen to align along the line-of-sight.

Since maser action only takes place when the gas density and temperature fall within the narrow ranges shown in Section \ref{sec:3.1}, one could assume that, on average, $n_{\rm u}\Delta P$ in the high-z environment would be comparable to that in local maser sources. Given this assumption and considering the optimal gain length $L_{\rm g} \approx L_{\rm c}$, the key parameters that determine the maser flux density would be the coherence path length $L_{\rm c}$ of the disk and the obvious inverse square dependence of the luminosity distance $D_{\rm L}$. 

For the high-velocity maser components in a maser disk, the velocity coherent path length $L_{\rm c}$ at the tangent points at radius $R$ is $L_{\rm c}=2(\delta V/V)^{1/2}R$, where $\delta V$ and $V$ are the gas velocity dispersion and the orbital velocity, respectively, suggesting that the gain length would increase linearly with $R$ \citep{lo05}. If the radius of a maser disk were increased by a factor of $\gtrsim 20-30$, the isotropic luminosity density of the source $L_{\nu}\equiv 4\pi D_{\rm L}^{2}S_{\nu}/(1+z)$ would increase by a factor of $\gtrsim$$8000-27000$ due to the increase in $L_{\rm c}$, making the source an H$_{2}$O gigamaser. 

Assuming such H$_{2}$O gigamaser disks exist at $z\sim 2-3$, their luminosity distances would be $\sim 150-250$ times greater than a maser disk at the distance of $\sim$100 Mpc assuming standard cosmology. The inverse square dependence of $D_{\rm L}$ would then lead to a decrease in the flux density by a factor of $\sim 20000-60000$. By considering the change in the flux density $S_{\nu}$ due to the factor $(1+z)$ and the increases in $D_{\rm L}$ and $L_{\rm c}$ based on Equation \ref{eq:maser_flux}, one would expect that the flux densities of a 22 GHz H$_{2}$O gigamasers at $z\sim 2-3$ could be comparable to that of a local H$_{2}$O megamaser at $\sim$100 Mpc. Given that the flux densities of known H$_{2}$O maser disks at $D_{\rm L}\sim 100$ Mpc are typically $\sim$$20-40$ mJy \citep[e.g.][]{kuo13, kuo15, gao16}, the expected flux densities of the strongest maser features in a high-z gigamaser could range from a few mJy up to $\gtrsim$40 mJy, suggesting that a few $\sigma$ detections of the strongest lines are possible with a few hours of on-source integration time using the VLA, the GBT, and the High Sensitivity Array (HSA). Note that before applying the above estimation to high-z sources, one needs to explore whether the high-z gigamaser disks could have significantly larger sizes than the ones in the local universe.

\subsubsection{The Dependence of Maser Disk Size on the Black Hole Mass}
Based on the observations of quasars in \citet{vo09}, it has been shown that the BH mass function of luminous quasars at redshifts $1.5 \lesssim z \lesssim 3$ peaks at $M_{\rm BH}\sim 1.5\times10^{9} M_{\odot}$, and the Eddington ratios of these quasars tend to be $\lambda_{\rm Edd}\gtrsim 0.1$. To see whether maser disks could exist around these $\gtrsim 10^{9} M_{\odot}$ supermassive black holes, with their sizes significantly larger than the low redshift maser systems, we explore the dependence of maser disk size on black hole mass based on the disk model presented in Section \ref{sec:3.4}. In this exploration, we adopt the Mestel disk profile $\Sigma(r) = \Sigma_{\rm out}(r/a_{\rm 0})^{-1}$ with $c_{\rm g}=2$ km~s$^{-1}$, $a_{0}=$ 1 pc, and fix the obliquity parameter at $\eta=0.15$, a typical value for local H$_{2}$O maser disks. In addition, we set the bolometric luminosity in our model as $L_{\rm bol}=\lambda_{\rm Edd}L_{\rm Edd}$, with $\lambda_{\rm Edd}$ varied between 0.03 to 0.6. Finally, we consider five representative values for the disk-mass-to-BH-mass ratio $\tilde{M}_{\rm D}/M_{\rm BH}$ that ranges from 0.001 to 0.03. Given a combination of $\lambda_{\rm Edd}$ and $\tilde{M}_{\rm D}/M_{\rm BH}$, we model the heating rate and density distribution in the disk and calculate the outer radius of the masing region for BH mass between $1.0\times 10^{7}M_{\odot}$ and $ 1.5\times 10^{9} M_{\odot}$. 

In the left panel of Figure \ref{fig:8}, we show $R_{\rm out}$ as a function of $M_{\rm BH}$ for $\lambda_{\rm Edd}$ = 0.03, 0.06, 0.1, 0.3, and 0.6, with $\tilde{M}_{\rm D}/M_{\rm BH} = 0.005$, which is $\sim$3 times greater than the mean disk-to-BH mass ratio for our sample of low-z maser disks. The right panel of Figure \ref{fig:8} shows the prediction of the outer radius for $\tilde{M}_{\rm D}/M_{\rm BH}$ ranging from 0.001 to 0.03, assuming $\lambda_{\rm Edd}=0.1$. It can be seen in both plots that the outer radius $R_{\rm out}$ for the cases with $M_{\rm D}/M_{\rm BH}\lesssim 0.005$ first increases with BH mass with a steeper slope, and the slope appears to drop distinctly when the BH mass is greater than a {\it critical} value that varies depending on the combination of $M_{\rm BH}$, $\lambda_{\rm Edd}$, and $\tilde{M}_{\rm D}$. We note that this critical value is simply the critical BH mass  $M_{\rm BH}^{\rm crit}$ discussed in Section \ref{sec:3.6}. The slope changes at $M_{\rm BH}\sim M_{\rm crit}$ because the maser disk transitions from the heating rate limited regime (i.e. $M_{\rm BH} << M_{\rm BH}^{\rm crit}$ and $R_{\rm out}^{\rm H} \propto M_{\rm BH}$) to the minimum density limited regime (i.e. $M_{\rm BH} >> M_{\rm BH}^{\rm crit}$ and $R_{\rm out}^{\rm D} \propto M_{\rm BH}^{0.6}$). 

As indicated by Equation \ref{eq:13}, the critical BH mass would be $M_{\rm BH}^{\rm crit}\lesssim 10^{8} M_{\odot}$ if $0.1 \lesssim \lambda_{\rm Edd} \lesssim 1$ and $\tilde{M}_{\rm D}/M_{\rm BH} \gtrsim 0.005$, suggesting that the outer radius of the maser disk around a $\gtrsim 10^{9} M_{\odot}$ BH in a high-z quasar would be determined by $R_{\rm out}^{\rm D}$ if the gas disk is massive enough. As one can see in Figure \ref{fig:8}, such high-z maser disks would have $R_{\rm out}\sim 10-30$ pc given $\tilde{M}_{\rm D}/M_{\rm BH}\sim 0.005 - 0.03$, about $\sim 20-60$ times greater than the average outer radius of the local maser disks (i.e. $\overline{R}_{\rm out} = 0.52$ pc). Based on the discussion in Section \ref{sec:4.3.1}, we speculate that such large disk sizes could lead to H$_{2}$O gigamasers in high redshift galaxies.

Considering the typical angular-diameter distance $D_{\rm A}$ for a galaxy at $z\sim 2-3$ (i.e. $D_{\rm A}\sim1620-1760$ Mpc), the angular radius of a high-z maser disk with the physical size of $r\sim 10-30$ pc would be $\sim1.2 - 3.8$ milliarcseconds, comparable to those of the local H$_{2}$O maser disks, suggesting that it is possible to apply the H$_{2}$O maser technique to high-z quasars with existing centimeter VLBI facilities. If one could further detect submillimeter water maser emissions \citep[e.g.][]{dom16,dom23} from these high-z H$_{2}$O gigamaser disks, it is possible that future observations with the Event Horizon Telescope (EHT) at submillimeter wavelengths would provide highly accurate maser imaging with $\sim20-40$ microarcsecond resolution at submillimeter wavelengths, leading to maser maps with fractional position uncertainties comparable to NGC 4258 \citep[e.g.][]{argon07, hum13}. Depending on the redshift and the specific transition, these high-redshift submillimeter masers may even be observable with the Global Millimeter VLBI Array (GMVA) at millimeter wavelengths. For example, a 321 GHz, 325 GHz, and 380 GHz maser \citep{gray16} at a redshift of $z=3$ would have an observed frequency of $\sim$80 GHz, $\sim$81 GHz,and $\sim$95 GHz, respectively, allowing for observations of these maser transitions with the 3mm receiver of the GMVA.

\section{Conclusion} \label{sec:5}

% -------------- Version 1 ---------------

In this work, we combine ideas from multiple previous studies and develop a warped molecular disk model to examine the distributions of gas density and X-ray heating rate in the X-ray illuminated disk, with the gas disk described by  
a power-law surface density profile $\Sigma(r) \propto r^{s}$. This allows us to identify the boundaries of the masing region within which both gas density and temperature fall in the favored ranges for maser excitation. We compare the observed radii of sixteen maser disks with the predictions from our model, with the main conclusions summarized as follows : 

\begin{itemize} 
%\item[1.] The predictions from the well-known NM95 model that assumes steady-state accretion tend to over-predict the outer radii of the maser disks by a factor of $\sim$$3-10$ for $\sim$$75\%$ of our sample if one approximates the mass accretion rate as $\dot{M}\approx L_{\rm bol}/\epsilon c^{2}$. In light of the results from our modeling, it is most likely that this discrepancy originates from the breakdown of the steady-state assumption for the majority of the maser disks. \dom{I still don't understand how we have determined that the steady-state assumption is the piece of the model that's deficient here.}

%we examine the distributions of gas density and X-ray heating rate in a warped molecular disk described by a power-law surface density profile. With a suitable choice of the disk mass, we find that the outer radius $R_{\rm out}$ of the maser disk predicted from our model can match the observed value, with $R_{\rm out}$ mainly determined by the maximum heating rate or the minimum density for efficient maser action, depending on the combination of the Eddington ratio, black hole mass and disk mass. 

\item[1.] With a suitable choice of disk mass, the predictions from our model agree reasonably well with the observed outer maser disk sizes from observations for all sixteen maser disks in our sample, with the agreement substantially better than previous models.

%The outer radii of all maser disks can be well explained if one adopts the disk model described by the power-law surface density profile. By examining the distributions of the X-ray heating rate and gas density in an X-ray illuminated molecular disk based on the power-law model, we are able identify the masing region within which the physical conditions of the gas would enable efficient maser action. For all maser disks in our sample, we can find solutions that predict disk outer radii consistent with the observations. The best-fit models reveal that the masing regions tend to lie at the mid-plane of the disk, with the outer boundaries defined either by the maximum X-ray heating rate or minimum gas density for maser pumping, depending on the combination of $M_{\rm BH}$, $M_{\rm D}$, and $\lambda_{\rm Edd}$. 

\item[2.] The outer edges of the maser disks are set by either the maximum X-ray heating or by the minimum gas density required for population inversion, depending on the critical black hole mass $M_{\rm BH}^{\rm crit}$ for a given system.

\item[3.] The critical BH mass $M_{\rm BH}^{\rm crit}$ is determined by the combination of $\lambda_{\rm Edd}$, $M_{\rm BH}$, $\tilde{M}_{\rm D}$, and $\eta$ of a maser system, and it separates the maser systems into the heating rate limited regime and the minimum density limited regime. This understanding explains two otherwise disparate previous results from \citet[][$R_{\rm out} \propto M_{\rm BH}$]{wy12} and from \citet[][$R_{\rm out} \propto M_{\rm BH}^{0.57\pm 0.16}$]{gao17}, because it predicts two different scaling laws depending on which regime a maser system falls into. 

%\item[3.] The physical conditions of the gas alone cannot explain the inner radii of the maser disks. In the region well inside the inner radius of a maser disk, one can always find gas at a sufficiently high elevation having physical conditions suitable for maser excitation, suggesting that the inner edge of a maser disk involves physics beyond basic gas properties.

\item[4.] The observed inner radii of the majority of the maser disks are roughly consistent with the dust sublimation radius, suggesting that $R_{\rm in}$ of a maser disk is probably set by dust sublimation rather than by disk warping effects as proposed by NM95, though NGC 4258 appears to be a significant outlier, presenting a potential counterexample that requires more investigation.

%We find that the observed inner radii of the majority of the maser disks are roughly consistent with the dust sublimation radius $R_{\rm sub, Nenkova}$ prescribed by \citet{nenkova08}, which indicates the transition radius between dusty and dust-free environments. It is likely that the trapping of far-infrared photons by the masing clouds becomes more significant as dust gradually sublimates away at this transition region, leading the inner edge of a maser disk where the population inversion is mostly quenched.  

\item[5.] By applying our model to high-z quasars, our work predicts that H$_{2}$O {\it gigamasers} should exist in the early universe, with their maser flux densities detectable with existing radio facilities.

%Finally, our model predicts that H$_{2}$O gigamaser disks could exist around $\gtrsim 10^{9} M_{\odot}$ supermassive BHs at the centers of high-z quasars. Their sizes could be as large as $\sim10-30$ pc if the disk-to-BH-mass ratio is comparable to or greater than the average value for local maser disks. The predicted flux densities of these systems range from a few mJy to $\gtrsim 20-30$ mJy, high enough to be detected with existing radio interferometers with a few hours of on-source integration. Future surveys of H$_{2}$O gigamasers from high-z quasars that could host these systems \citep[e.g. Compton-thick AGNs;][]{kuo20b} would provide a good test for our model.

\end{itemize}

% ---------- Verion 2 ------------------
%In this work, we examine whether the physical conditions favorable for population inversion of H$_{2}$O molecules can be the primary factors that determine radii of the inner and outer edges of the sixteen H$_{2}$O megamaser disks that have high-qaulity VLBI maps. We first test the predictions from the well-known NM95 model, followed by considering a non-steady-state disk model in which the disk surface density described by the power-law profile. By assuminig that maser action takes place in the X-ray dissociation region in a warped molecular disk irradiated by the central X-ray source, we calculate X-ray heating rate in the disk and identify the masing region in the disk within which the gas density and temperature would fall within the suitable ranges that enable maser action. Our conclusions are summarized as follows : 

%\begin{itemize} 
%\item[1.] 
 
%\item[2.] 
%\item[3.] 
%\item[4.] Finally, our model predicts that H$_{2}$O {\it gigamaser} disks could exist around $\gtrsim 10^{9} M_{\odot}$ supermassive BHs at the centers of high-z quasars, and their sizes could be as large as $\sim10-30$ pc. The predicted flux densities of these systems range from a few mJy to $\gtrsim 20-30$ mJy, high enough to be detected with exiting radio interferometers and VLBI facilities with a few hours of on-source intergration.

%\end{itemize} 

\section*{Acknowledgements}

We gratefully thank Dr. Fred Lo, the former director of National Radio Astronomy Observatory, for initiating the work for this paper before he passed away in 2016. This publication is supported by Ministry of Science and Technology, R.O.C. under the project 112-2112-M-110-003. This research has made use of NASA's Astrophysics Data System Bibliographic Services, and the NASA/IPAC Extragalactic Database (NED) which is operated by the Jet Propulsion Laboratory, California Institute of
Technology, under contract with the National Aeronautics and Space Administration. In addition, this work also makes use of the cosmological calculator described in \citet{wright06}.

%%%%%%%%%%%%%%%%%%%%%%%%%%%%%%%%%%%%%%%%%%%%%%%%%%
\section*{Data Availability}

The data underlying this article are available in the article and in its online supplementary material. 
%The inclusion of a Data Availability Statement is a requirement for articles published in MNRAS. Data Availability Statements provide a standardised format for readers to understand the availability of data underlying the research results described in the article. The statement may refer to original data generated in the course of the study or to third-party data analysed in the article. The statement should describe and provide means of access, where possible, by linking to the data or providing the required accession numbers for the relevant databases or DOIs.



\bibliographystyle{mnras}
\bibliography{references} % if your bibtex file is called example.bib

%%%%%%%%%%%%%%%%%%%%%%%%%%%%%%%%%%%%%%%%%%%%%%%%%%

%%%%%%%%%%%%%%%%% APPENDICES %%%%%%%%%%%%%%%%%%%%%

%\appendix



%%%%%%%%%%%%%%%%%%%%%%%%%%%%%%%%%%%%%%%%%%%%%%%%%%

% Don't change these lines
%\bsp	% typesetting comment
\label{lastpage}
\end{document}